\documentclass[twoside, 11pt]{article}

\newif\ifjmlr
\jmlrfalse

\usepackage[ruled]{algorithm2e}

\usepackage[margin=1in]{geometry}

\usepackage{amsmath}
\usepackage{amssymb}
\usepackage{amsfonts}
\usepackage{epsfig}
\usepackage{array}
\usepackage{subfig}
\usepackage{times}
\usepackage{appendix}
\usepackage{enumitem}
\usepackage{comment}

\ifjmlr
\usepackage{jmlr2e} 
\jmlrheading{1}{2000}{1-48}{4/00}{10/00}{Shah and Zhou}
\ShortHeadings{Double or Nothing}{Shah and Zhou}
\let\chapter\section
\fi

\makeatletter
\newenvironment{subtheorem}[1]{%
  \def\subtheoremcounter{#1}%
  \refstepcounter{#1}%
  \protected@edef\theparentnumber{\csname the#1\endcsname}%
  \setcounter{parentnumber}{\value{#1}}%
  \setcounter{#1}{0}%
  \expandafter\def\csname the#1\endcsname{\theparentnumber.\Alph{#1}}%
  \ignorespaces
}{%
  \setcounter{\subtheoremcounter}{\value{parentnumber}}%
  \ignorespacesafterend
}
\makeatother
\newcounter{parentnumber}

\usepackage[hidelinks]{hyperref}

\ifjmlr
	\usepackage{natbib}
	\newcommand{\citetn}{\citet}
	\newcommand{\citepn}{\citep}
\else
	\newcommand{\citetn}{\cite}
	\newcommand{\citepn}{\cite}
\fi

\ifjmlr
\else
	 
	\newtheorem{theorem}{Theorem}
	\newtheorem{axiom}[theorem]{Axiom}
	\newtheorem{proposition}[theorem]{Proposition}
	\newtheorem{lemma}[theorem]{Lemma}
	\newtheorem{corollary}[theorem]{Corollary}
	\newtheorem{remark}{Remark}
	\newtheorem{example}{Example}
\fi

\newcommand{\beq}{\begin{equation}}
\newcommand{\eeq}{\end{equation}}
\newcommand{\bea}{\begin{eqnarray}}
\newcommand{\eea}{\end{eqnarray}}
\newcommand{\bean}{\begin{eqnarray*}}
\newcommand{\eean}{\end{eqnarray*}}
\newcommand{\bit}{\begin{itemize}}
\newcommand{\eit}{\end{itemize}}
\newcommand{\ben}{\begin{enumerate}}
\newcommand{\een}{\end{enumerate}}
\newcommand{\blem}{\begin{lem}}
\newcommand{\elem}{\end{lem}}
\newcommand{\bthm}{\begin{thm}}
\newcommand{\ethm}{\end{thm}}

\newcommand{\elll}{l}
\newcommand{\numques}{N}
\newcommand{\numgold}{G}
\newcommand{\ans}[1]{{x_{#1}}}
\newcommand{\ansbf}{\boldsymbol{x}}
\newcommand{\payfn}{f}
\newcommand{\maxpay}{\mu_{\max}}
\newcommand{\minpay}{\mu_{\min}}

\newenvironment{myquote}{\list{}{\leftmargin=0.3in\rightmargin=0.3in}\item[]}{\endlist}

\title{Double or Nothing:\\ Multiplicative Incentive Mechanisms for Crowdsourcing}

\ifjmlr
	\author{\name Nihar B. Shah \email nihar@eecs.berkeley.edu\\ 
	\addr Department of Electrical Engineering and Computer Sciences\\ 
	University of California, Berkeley\\
	Berkeley, CA 94720 USA
	\AND 
	\name Dengyong Zhou \email dengyong.zhou@microsoft.com\\ 
	\addr Machine Learning Department\\ 
	Microsoft Research\\
	One Microsoft Way, Redmond 98052 USA} 	
	\editor{}
\else
	\author{\begin{tabular}{ccc}Nihar B. Shah &\quad& Dengyong Zhou\\Department of EECS
	& \quad & 	
	Machine Learning Department\\University of California, Berkeley & \quad & Microsoft Research\\ {\tt nihar@eecs.berkeley.edu}& & {\tt dengyong.zhou@microsoft.com}\end{tabular}
	}	
    \date{}
\fi
\begin{document}

\maketitle

\begin{abstract}%
Crowdsourcing has gained immense popularity in machine learning applications for obtaining large amounts of labeled data. Crowdsourcing is cheap and fast, but suffers from the problem of low-quality data. To address this fundamental challenge in crowdsourcing, we propose a simple payment mechanism to incentivize workers to answer only the questions that they are sure of and skip the rest. We show that surprisingly, under a mild and natural ``no-free-lunch'' requirement, this mechanism is the one and only incentive-compatible payment mechanism possible. We also show that among all possible incentive-compatible  mechanisms (that may or may not satisfy no-free-lunch), our mechanism makes the smallest possible payment to spammers. We further extend our results to a more general setting in which workers are required to provide a quantized confidence for each question.  Interestingly, this unique mechanism takes a ``multiplicative'' form. The simplicity of the mechanism is an added benefit. In preliminary experiments involving over 900 worker-task pairs, we observe a significant drop in the error rates under this unique mechanism for the same or lower monetary expenditure.
\end{abstract}

\ifjmlr
\fi

\section{Introduction}\label{sec:intro}

Complex machine learning tools such as deep learning are gaining increasing popularity and are being applied to a wide variety of problems. These tools require large amounts of labeled data~\citepn{hinton2012deep,raykar2010learning, deng2009imagenet, carlson2010coupled}.  These large labeling tasks are being performed by coordinating crowds of semi-skilled workers through the Internet. This is known as crowdsourcing. Generating large labeled data sets through crowdsourcing is inexpensive and fast as compared to employing experts. Furthermore, given the current platforms for crowdsourcing such as Amazon Mechanical Turk and many others, the initial overhead of setting up a crowdsourcing task is minimal. Crowdsourcing as a means of collecting labeled training data has now become indispensable to the engineering of intelligent systems.  The crowdsourcing of labels is also often used to supplement automated algorithms, to perform the tasks that are too difficult to accomplish by machines alone~\citepn{khatib2011crystal,lang2011using,bernstein2010soylent,von2008recaptcha,franklin2011crowddb}.

Most workers in crowdsourcing are not experts. As a consequence, labels obtained from crowdsourcing typically have a significant amount of error~\citepn{kazai2011crowdsourcing, vuurens2011much,wais2010towards}. It is not surprising that there is significant emphasis on having higher quality labeled data for machine learning algorithms, since a higher amount of noise implies requirement of more labels for obtaining the same accuracy in practice. Moreover, several algorithms and settings are not very tolerant of data that is noisy~\citepn{long2010random, hanneke2010negative, manwani2013noise, baldridge2009well}; for instance,~\citetn{long2010random} conclude that ``a range of different types of boosting algorithms that optimize a
convex potential function satisfying mild conditions cannot tolerate random classification noise.'' Recent efforts have focused on developing statistical techniques to post-process the noisy labels in order to improve its quality (e.g.,~\citetn{raykar2010learning, zhou2012learning,wauthier2011bayesian, chen2013pairwise,dawid1979maximum, karger2011iterative,liu2012variational, zhang2014spectral, ipeirotis2014repeated}). However, when the inputs to these algorithms are very erroneous, it is difficult to guarantee that the processed labels will be reliable enough for subsequent use by machine learning or other applications. In order to avoid ``garbage in, garbage out'', we take a complementary approach to this problem: cleaning the data at the time of collection.

\begin{figure}
\centering
\includegraphics[width=.7\textwidth]{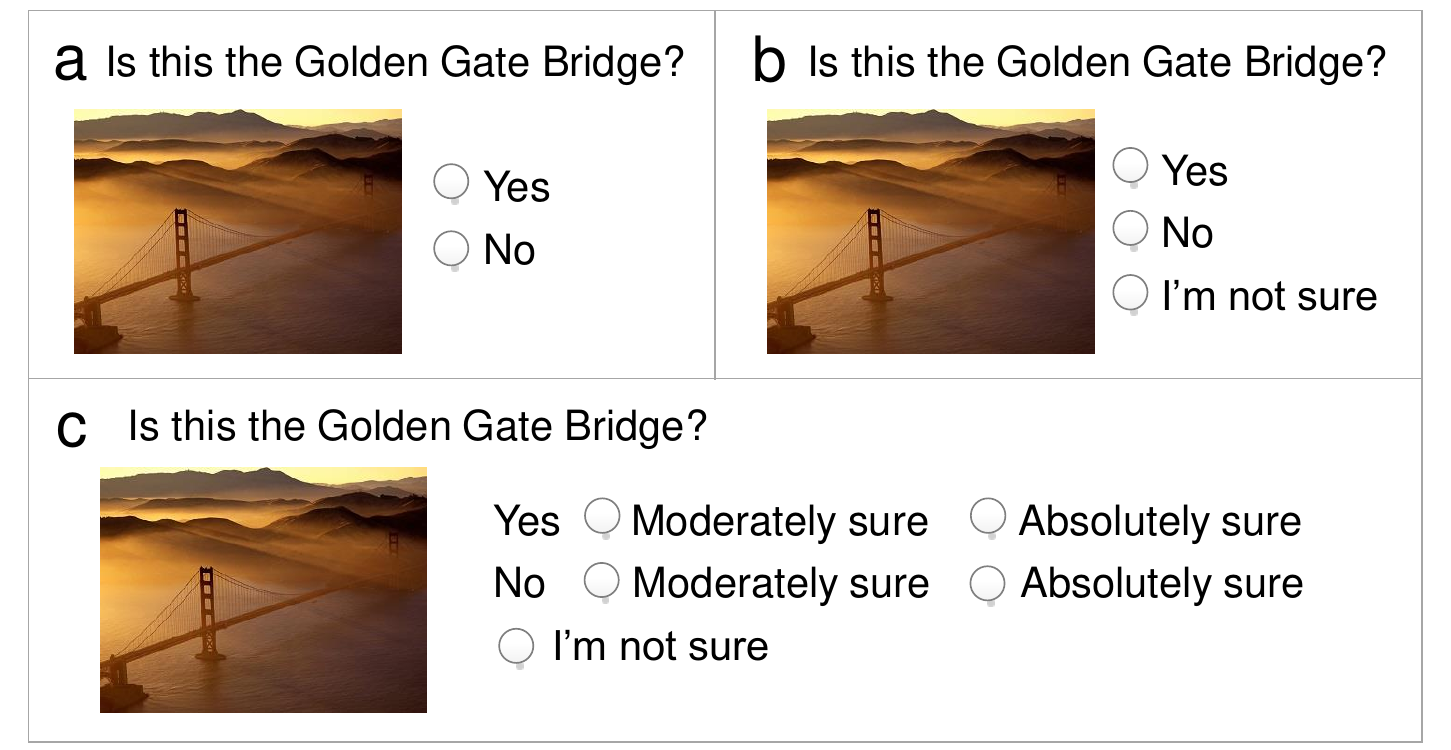}
\caption{Different interfaces for a task that requires the worker to answer the question ``Is this the Golden Gate Bridge?'': (a) the conventional interface; (b) with an option to skip; (c) with multiple confidence levels.}
\label{fig:intro}
\end{figure}

We consider crowdsourcing settings where the workers are paid for their services, such as in the popular crowdsourcing platforms of Amazon Mechanical Turk (\url{mturk.com}), Crowdflower (\url{crowdflower.com}) and other commercial platforms, as well as internal crowdsourcing platforms of companies such as Google, Facebook and Microsoft. These commercial platforms have gained substantial popularity due to their support for a diverse range of tasks for machine learning labeling, varying from image annotation and text recognition to speech captioning and machine translation.  We consider problems that are objective in nature, that is, have a definite answer. Figure~\ref{fig:intro}a depicts an example of such a question where the worker is shown a set of images, and for each image, the worker is required to identify if the image depicts the Golden Gate Bridge.

Our approach builds on the simple insight that in typical crowdsourcing setups, workers are simply paid in proportion to the amount of tasks they complete. As a result, workers attempt to answer questions that they are not sure of, thereby increasing the error rate of the labels.  For the questions that a worker is not sure of, her answers could be very unreliable~\citepn{wais2010towards,kazai2011crowdsourcing, vuurens2011much,jagabathula2014reputation}. To ensure acquisition of only high-quality labels, we wish to encourage the worker to skip the questions about which she is unsure, for instance, by providing an explicit ``I'm not sure'' option for every question (see Figure~\ref{fig:intro}b).  Given this additional option, one must also ensure that the worker is indeed incentivized to skip the questions that she is not confident about.  In a more general form, we consider eliciting the confidence of the worker for each question at multiple levels. For instance, in addition to ``I'm not sure'', we may also provide options like ``absolutely sure'', and ``moderately sure'' (see Figure~\ref{fig:intro}c). The goal is to design payment mechanisms that incentivize the worker to attempt only those questions for which they are confident enough, or alternatively, report their confidences truthfully. As we will see later, this significantly improves the aggregate quality of the labels that are input to the machine learning algorithms. We will term any payment mechanism that incentivizes the worker to do so as ``incentive compatible''.

In addition to incentive compatibility, preventing spammers is another desirable requirement from incentive mechanisms in crowdsourcing. Spammers are workers who answer randomly without regard to the question being asked, in the hope of earning some free money, and are known to exist in large numbers on crowdsourcing platforms~\citepn{wais2010towards,bohannon2011social, kazai2011crowdsourcing, vuurens2011much}. The presence of spammers can significantly affect the performance of any machine learning algorithm that is trained on this data.
It is thus of interest to deter spammers by paying them as low as possible. An intuitive objective, to this end, is to ensure a minimum possible expenditure on spammers who answer randomly. For instance, in a task with binary-choice questions, a spammer is expected to have half of the attempted answers incorrect; one may thus wish to set the payment to its minimum possible value if half or more of the attempted answers are wrong.
In this paper, however, we impose \emph{strictly and significantly weaker requirement}, and then show that there is one and only one incentive-compatible mechanism that can satisfy this weak requirement. Our requirement is referred to as the ``no-free-lunch'' axiom. In the skip-based setting, it says that if \emph{all} the questions attempted by the worker are answered incorrectly, then the payment must be the minimum possible. The no-free-lunch axiom for the general confidence-based setting is even weaker: if the worker indicates the highest confidence level for \emph{all} the questions she attempts in the gold standard, and furthermore if all these responses are incorrect, then the payment must be the minimum possible. We term this condition the ``no-free-lunch'' axiom. In the general confidence-based setting, we want to make the minimum possible payment if the worker indicates the \emph{highest confidence level} for \emph{all} the questions she attempts \emph{and} if \emph{all} these responses are incorrect.

In order to test whether our mechanism is practically viable, and to assess the quality of the final labels obtained, we conducted experiments on the Amazon Mechanical Turk crowdsourcing platform. In our preliminary experiments that involved several hundred workers, we found that the quality of data consistently improved by use of our schemes as compared to the standard settings, often by two-fold or higher, with the total monetary expenditure being the same or lower as compared to the conventional baseline.

\ifjmlr
\subsection{Summary of Contributions}
\else
\paragraph{Summary of Contributions.}
\fi
 We propose a payment mechanism for the aforementioned setting (``incentive compatibility'' plus ``no-free-lunch"), and show that surprisingly, this is the \emph{only} possible mechanism. We also show that additionally, our mechanism makes the smallest possible payment to spammers among all possible incentive compatible mechanisms that may or may not satisfy the no-free-lunch axiom. Interestingly, our payment mechanism takes a multiplicative form: the evaluation of the worker's response to each question is a certain score, and the final payment is a \emph{product} of these scores. This mechanism has additional appealing features in that it is simple to compute, and is also simple to explain to the workers. Our mechanism is applicable to any type of objective questions, including multiple choice annotation questions, transcription tasks, etc.  In preliminary experiments on Amazon Mechanical Turk involving over 900 worker-task pairs, the quality of data improved significantly under our unique mechanism, with the total monetary expenditure being the same or lower as compared to the conventional baseline.

\ifjmlr
    \subsection{Related Literature}
\else
    \paragraph{Related Literature.}
\fi
The framework of ``strictly proper scoring rules''~\citepn{brier1950verification,savage1971elicitation, gneiting2007strictly,lambert2009eliciting} provides a general theory for eliciting information for settings where this information can subsequently be verified by the mechanism designer, for example, by observing the true value some time in the future. In our work, this verification is performed via the presence of some ``gold standard'' questions in the task. Consequently, our mechanisms can also be called ``strictly proper scoring rules''. It is important to note that the framework of strictly proper scoring rules, however, provides a large collection of possible mechanisms and does not guide the choice of a specific mechanism from this collection~\citepn{gneiting2007strictly}. In this work, we show that for the crowdsourcing setups considered, under a very mild ``no-free-lunch'' condition, the mechanism proposed in this paper is the one and only strictly proper scoring rule.

Interestingly, proper scoring rules have another interesting connection with machine learning techniques: to quote~\citetn{buja2005loss}, ``proper scoring rules comprise most loss functions currently in use: log-loss,
squared error loss, boosting loss, and as limiting cases cost-weighted misclassification losses.'' The present paper does not investigate this aspect of  proper scoring rules, and we refer the reader to~\citetn{buhlmann2007boosting, mease2007boosted, buja2005loss} for more details.


The design of statistical inference algorithms for denoising the data obtained from workers is an active topic of research~\citepn{raykar2010learning,zhou2012learning,wauthier2011bayesian, chen2013pairwise,kamar2012combining,dawid1979maximum, karger2011iterative,liu2012variational,zhang2014spectral, vempaty2014reliable, ipeirotis2014repeated}. In addition, several machine learning algorithms accommodating errors in the data have also been designed~\citepn{angluin1988learning,cano2001learning,lee2004hybrid,chu2004adaptive}. These algorithms are typically oblivious to the elicitation procedure. Our work nicely complements this line of research in that these inference algorithms may now additionally employ the higher quality data and the specific structure of the elicited data for an improved denoising efficiency.

Another relevant problem in crowdourcing is that of choosing which workers to hire or efficiently matching workers to tasks, and such problems are studied in~\citetn{yuen2011task,ho2013adaptive,zhou2014optimal,anari2014mechanism} under different contexts. Our work assumes that a worker is already matched, and focuses on incentivizing that worker to respond in a certain manner. A recent line of work has focussed on elicitation of data from multiple agents in order to perform certain specific estimation tasks~\citepn{fang2007putting,dekel2008incentive,cai2014optimum}. In contrast, our goal is to ensure that workers censor their own low-quality (raw) data, without restricting our attention to any specific downstream algorithm or task. 


\paragraph{Organization.} The organization of this paper is as follows. We present the formal problem setting in Section~\ref{sec:setting}. In Section~\ref{sec:skip} we consider the skip-based setting: We present our proposed mechanism and show that it is the only mechanism which satisfies the requirements discussed above. In Section~\ref{sec:confidence}, we then consider the more general setting of eliciting a quantized value of the worker's confidence. We construct a mechanism for this setting, which also takes a multiplicative form, and prove its uniqueness. In Section~\ref{sec:strong_skip} we prove that imposing a requirement that is only slightly stronger than our proposed no-free-lunch axiom leads to impossibility results. In Section~\ref{sec:sim_exp} we present synthetic simulations and real-world experiments on Amazon Mechanical Turk to evaluate the potential of our setting and algorithm to work in practice. We conclude the paper with a discussion on the various modeling choices, future work, and concluding remarks in Section~\ref{sec:conclusions}. 

The paper contains three appendices. In Appendix~\ref{sec:proofs} we prove all theoretical results whose proofs are not presented in the main text. We provide more details of the experiments in Appendix~\ref{app:experiments}. In Appendix~\ref{sec:utility} we extend our results to a setting where workers aim to maximize the expected value of some ``utility'' of their payments. 

\section{Setting and Notation}\label{sec:setting}
In the crowdsourcing setting that we consider, one or more workers perform a \textit{task}, where a task consists of multiple \textit{questions}. The questions are objective, by which we mean, each question has precisely one correct answer. Examples of objective questions include multiple-choice classification questions such as Figure~\ref{fig:intro}, questions on transcribing text from audio or images, etc.

For any possible answer to any question, we define the worker's \textit{confidence about an answer} as the probability, according to her belief, of this answer being correct. In other words, one can assume that the worker has (in her mind) a probability distribution over all possible answers to a question, and the confidence for an answer is the probability of that answer being correct. As a shorthand, we also define the \textit{confidence about a question} as the confidence for the answer that the worker is most confident about for that question. We assume that the worker's confidences for different questions are independent. Our goal is that for every question, the worker should be incentivized to skip if her confidence for that question is below a certain pre-defined threshold, otherwise select the answer that she is most confident about, and if asked, also indicate a correct (quantized) value of her confidence for the answer.

Specifically, we consider two settings:
\begin{itemize}[leftmargin=*]
\item \textbf{Skip-based}.  For each question, the worker can either choose to `skip' the question or provide an answer (Figure~\ref{fig:intro}b).
\item \textbf{Confidence-based}. For each question, the worker can either `skip' the question or provide an answer, and in the latter case, indicate her confidence for this answer as a number in $\{1,\ldots,L\}$  (Figure~\ref{fig:intro}c). We term this indicated confidence as the `confidence-level'. Here, $L$ represents the highest confidence-level, and `skip' is considered to be a confidence-level of $0$.~\footnote{When the task is presented to the workers, the word `skip' or the numbers $\{1,\ldots,L\}$ are replaced by more comprehensible phrases such as ``I don't know'', ``moderately sure'', ``absolutely sure'', etc.}
\end{itemize}
One can see from the aforementioned definition that the confidence-based setting is a generalization of the skip-based setting (the skip-based setting corresponds to $L=1$). The goal is to ensure that for a given set of intervals that partition $[0,1]$, for every question the worker is incentivized to indicate `skip' or choose the appropriate confidence-level when her confidence for that question falls in the corresponding interval. The choice of these intervals will be defined subsequently in the skip-based and confidence-based sections (Section~\ref{sec:skip} and Section~\ref{sec:confidence}) respectively.

Let ${\numques}$ denote the total number of questions in the task. Among these questions, we assume the existence of some ``gold standard'' questions, that is, a set of questions whose answers are known to the requester. Let $\numgold~(1 \leq \numgold \leq {\numques})$ denote the number of gold standard questions.  The ${\numgold}$ gold standard questions are assumed to be distributed uniformly at random in the pool of ${\numques}$ questions (of course, the worker does not know which ${\numgold}$ of the ${\numques}$ questions form the gold standard). The payment to a worker for a task is computed after receiving her responses to all the questions in the task.  The payment is based on the worker's performance on the gold standard questions. Since the payment is based on known answers, the payments to different workers do not depend on each other, thereby allowing us to consider the presence of only one worker without any loss in generality.

We will employ the following standard notation. For any positive integer $K$, the set $\{1,\ldots,K\}$ is denoted by $[K]$. The indicator function is denoted by $\mathbf{1}$, i.e., $\mathbf{1}\{z\} = 1$ if $z$ is true, and $0$ otherwise. 


Let $\ans{1},\ldots,\ans{\numgold}$ denote the evaluations of the answers that the worker gives to the $\numgold$ gold standard questions, and let $\payfn$ denote the scoring rule, i.e., a function that determines the payment to the worker based on these evaluations $\ans{1},\ldots,\ans{\numgold}$. 

In the skip-based setting, $\ans{i} \in \{-1,0,+1\}$ for all $i \in [\numgold]$. Here, ``$0$'' denotes that the worker skipped the question, ``$-1$'' denotes that the worker attempted to answer the question and that answer was incorrect, and ``$+1$'' denotes that the worker attempted to answer the question and that answer was correct. The payment function is $\payfn:\{-1,0,+1\}^{\numgold} \rightarrow \mathbb{R}$.

In the confidence-based setting, $\ans{i} \in \{-L,\ldots,+L\}$ for all $i \in [\numgold]$. Here, we set $\ans{i}=0$ if the worker skipped the question, and for $\elll \in \{1,\ldots,L\}$,  we set $\ans{i}=\elll$ if the question was answered correctly with confidence $\elll$ and $\ans{i}=-\elll$ if the question was answered incorrectly with confidence $\elll$. The function $\payfn:\{-L,\ldots,+L\}^{\numgold} \rightarrow \mathbb{R}$ specifies the payment to be made to the worker.

The payment is further associated to two parameters, $\maxpay$ and $\minpay$. The parameter $\maxpay$ denotes the {\it budget}, i.e., the maximum amount that is paid to any individual worker for this task:
\begin{align*}
\max_{\ans{1},\ldots,\ans{\numgold}} \payfn(\ans{1},\ldots,\ans{\numgold}) = \maxpay.
\end{align*}
The amount $\maxpay$ is thus the amount of compensation paid to a perfect worker for her work. Further, one may often also have the requirement of paying a certain minimum amount to any worker. The parameter $\minpay~(\leq \maxpay)$ denotes this minimum payment: the payment function must also satisfy
\begin{align*}
\min_{\ans{1},\ldots,\ans{\numgold}} \payfn(\ans{1},\ldots,\ans{\numgold}) \geq \minpay.
\end{align*}
For instance, crowdsourcing platforms today allow payments to workers, but do not allow imposing penalties: this condition gives $\minpay = 0$. 



We assume that the worker attempts to maximize her overall expected payment. In what follows, the expression `the worker's expected payment' will refer to the expected payment from the worker's point of view, and the expectation will be taken with respect to the worker's confidences about her answers and the uniformly random choice of the ${\numgold}$ gold standard questions among the ${\numques}$ questions in the task. For any question $i \in [\numques]$, suppose the worker indicates the confidence-level $y_i \in \{0,\ldots,L\}$. Further, for every question $i \in [\numques]$ such that $y_i \neq 0$, let $p_i$ be the confidence of the worker for the answer she has selected for question $i$, and for every question $i \in [\numques]$ such that $y_i = 0$, let $p_i \in (0,1)$ be any arbitrary value. Let $E = (\epsilon_1,\ldots,\epsilon_\numgold) \in \{-1,1\}^\numgold$. Then from the worker's perspective, the expected payment for the selected answers and confidence-levels is
\begin{align}
\frac{1}{{{\numques} \choose {\numgold}}} \sum_{\substack{(j_1,...,j_{\numgold})\\\subseteq\{1,\ldots,{\numques}\}}} \sum_{E\in\{-1,1\}^{\numgold}} \left( \payfn( \epsilon_1 y_{j_1},\ldots,\epsilon_\numgold y_{j_\numgold} )  \prod_{i=1}^{\numgold} (p_{j_i})^{\frac{1+\epsilon_i}{2}} (1-p_{j_i})^{\frac{1-\epsilon_i}{2}} \right)~. \nonumber
\end{align}
In the expression above, the outermost summation corresponds to the expectation with respect to the randomness arising from the unknown positions of the gold standard questions. The inner summation corresponds to the expectation with respect to the worker's beliefs about the correctness of her responses.

A payment function $\payfn$ is called a \emph{incentive compatible} if the expected payment of the worker under this payment function is \emph{strictly} maximized when the worker answers in the manner desired. The specific requirements of the skip-based and the confidence-based settings are discussed subsequently in their respective sections.  We begin with the skip-based setting.

\section{Skip-based Setting}\label{sec:skip}
In this section, we consider the setting where for every question, the worker can choose to either answer the question or to skip it; no additional information is asked from the worker. See Figure~\ref{fig:intro}b for an illustration.

\subsection{Setting}\label{sec:skip_setting}
Let $T \in (0,1)$ be a predefined value. The goal is to design payment mechanisms that incentivize the worker to skip the questions for which her confidence is lower than $T$, and answer those for which her confidence is higher than $T$.~\footnote{In the event that the confidence about a question is exactly equal to $T$, the worker may choose to answer or skip.} Moreover, for the questions that she attempts to answer, she must be incentivized to select the answer that she believes is most likely to be correct. The value of $T$ is chosen apriori based on factors such as budget constraints or the targeted quality of labels. The value of $T$ may also depend on the choice of the algorithm that will subsequently be employed to aggregate the answers provided by multiple workers. In this paper, we will assume that the value of the threshold $T$ is already specified to us.

We impose the following simple and natural requirement:
\begin{axiom}[\bf No-free-lunch Axiom] If {all} the answers attempted by the worker in the gold standard are wrong, then the payment is the minimum possible. More formally, $\payfn(\ans{1},\ldots,\ans{\numgold})= \minpay$ for every evaluation $(\ans{1},\ldots,\ans{\numgold})$ such that $
0 < \sum_{i=1}^{\numgold} \mathbf{1}\{\ans{i} \neq 0\} = \sum_{i=1}^{\numgold} \mathbf{1}\{\ans{i} = -1\}$.
\end{axiom} 
One may expect a payment mechanism to impose the restriction of minimum payment to spammers who answer randomly. For instance, in a task with binary-choice questions, a spammer is expected to have $50\%$ of the attempted answers incorrect; one may thus wish to set a the minimum possible payment if $50\%$ or more of the attempted answers were incorrect. The no-free-lunch axiom which we impose is however a \emph{significantly weaker condition}, mandating minimum payment if \textit{all} attempted answers are incorrect.


\subsection{Payment Mechanism}
\label{subsec:payment}
We now present our proposed payment mechanism in Algorithm~\ref{algo:incentive_skip}.

\begin{algorithm}[H]
\begin{itemize}
\item Inputs:
\begin{itemize}
\item Threshold $T$
\item Budget parameters $\maxpay$ and $\minpay$
\item Evaluations $(\ans{1},\ldots,\ans{\numgold}) \in \{-1,0,+1\}^\numgold$ of the worker's answers to the $\numgold$ gold standard questions
\end{itemize}
\item Set $\alpha_{-1}=0,~\alpha_{0}=1,~\alpha_{+1}=\frac{1}{T}$
\item The payment is
\begin{equation*}
f(\ans{1},\ldots,\ans{\numgold}) = \kappa \prod_{i=1}^{\numgold}  \alpha_{\ans{i}} + \minpay,
\end{equation*}
where $\kappa = (\maxpay - \minpay) T^\numgold$.
\end{itemize}
\caption{Incentive mechanism for skip-based setting}
\label{algo:incentive_skip}
\end{algorithm}

The proposed mechanism has a  \emph{multiplicative} form: each answer in the gold standard is given a score based on whether it was correct (score = $\frac{1}{T}$), incorrect (score = $0$) or skipped (score = $1$), and the final payment is simply a product of these scores (scaled and shifted by constants). The mechanism is easy to describe to workers: For instance, if $T = \frac{1}{2}$, $\numgold=3$, $\maxpay=80$ cents and $\minpay=0$ cents, then the description reads:
\begin{myquote}{\it ``{The reward starts at 10 cents. For every correct answer in the 3 gold standard questions, the reward will double. However, if any of these questions are answered incorrectly, then the reward will become zero. So please use the `I'm not sure' option wisely.}''}\end{myquote} 
Observe how this payment rule is similar to the popular `double or nothing' paradigm~\citepn{doubleornothing}. 

The algorithm makes a minimum payment if \emph{one or more} attempted answers in the gold standard are wrong. Note that this property is significantly stronger than the property of no-free-lunch which we originally required, where we wanted a minimum payment only when \emph{all} attempted answers were wrong. Surprisingly, as we prove shortly, Algorithm~\ref{algo:incentive_skip} is the only incentive-compatible mechanism that satisfies no-free-lunch.

The following theorem shows that this mechanism indeed incentivizes a worker to skip the questions for which her confidence is below $T$, while answering those for which her confidence is greater than $T$. In the latter case, the worker is incentivized to select the answer which she thinks is most likely to be correct.
\begin{theorem}
The mechanism of Algorithm~\ref{algo:incentive_skip} is incentive-compatible and satisfies the no-free-lunch condition.
\label{thm:mechanism_working_skip}
\end{theorem}

In the remainder of this subsection, we present the proof of Theorem~\ref{thm:mechanism_working_skip}. The reader may go directly to subsection~\ref{subsec:skip_unique} without loss in continuity.

\paragraph{Proof of Theorem~\ref{thm:mechanism_working_skip}.} 
The proposed payment mechanism satisfies the no-free-lunch condition since the payment is $\minpay$ when there are one or more wrong answers in the gold standard. It remains to show that the mechanism is incentive compatible. To this end, observe that the property of incentive-compatibility does not change upon any shift of the mechanism by a constant value or any scaling by a positive constant value. As a result, for the purposes of this proof, we can assume without loss of generality that $\minpay=0$. 

We will first assume that, for every question that the worker does not skip, she selects the answer which she believes is most likely to be correct. Under this assumption we will show that the worker is incentivized to skip the questions for which her confidence is smaller than $T$ and attempt if it is greater than $T$. Finally, we will show that the mechanism indeed incentivizes the worker to select the answer which she believes is most likely to be correct for the questions that she doesn't skip. In what follows, we will employ the notation $\kappa = \maxpay T^\numgold$.

Let us first consider the case when ${\numgold}={\numques}$. Let $p_1,\ldots,p_{\numques}$ be the confidences of the worker for to questions $1,\ldots,{\numques}$ respectively. Further, let $p_{(1)}\geq \cdots \geq p_{(m)} > T > p_{(m+1)} \geq \cdots \geq p_{({\numques})} $ be the ordered permutation of these confidences (for some number $m$). Let $\{(1),\ldots,({\numques})\}$ denote the corresponding permutation of the ${\numques}$ questions. 
If the mechanism is incentive compatible, then the expected payment received by this worker should be maximized when the worker answers questions $(1),\ldots,(m)$ and skips the rest. Under the mechanism proposed in Algorithm~\ref{algo:incentive_skip}, this action fetches the worker an expected payment of
\[\kappa\frac{p_{(1)}}{T}\cdots\frac{p_{(m)}}{T}~.\] 
Alternatively, if the worker answers the questions $\{i_1,\ldots,i_z\}$, with $p_{i_1}<\cdots<p_{i_y} < T < p_{i_{y+1}}<\cdots p_{i_z}$, then the expected payment is 
\bea 
p_{i_1}\cdots p_{i_z}\frac{\kappa}{T^z}&=&\kappa\frac{p_{i_1}}{T}\cdots \frac{p_{i_z}}{T}\label{eq:mechanism_working_skip_ineq0}\\
&\leq &\kappa\frac{p_{i_1}}{T}\cdots \frac{p_{i_y}}{T}\label{eq:mechanism_working_skip_ineq1}\\
&\leq&\kappa\frac{p_{(1)}}{T}\cdots \frac{p_{(m)}}{T} \label{eq:mechanism_working_skip_ineq2}
\eea
where inequality~\eqref{eq:mechanism_working_skip_ineq1} holds because $\frac{p_{i_j}}{T} \leq 1~~\forall~j>y$ and holds with equality only when $z=y$. Inequality~\eqref{eq:mechanism_working_skip_ineq2} is a result of $\frac{p_{(j)}}{T} \geq 1~~\forall~j\leq m$ and holds with equality only when $y = m$. It follows that the expected payment is (strictly) maximized when $i_1=(1),\ldots,i_z=(m)$ as required.

The case of ${\numgold}<{\numques}$ is a direct consequence of the result for ${\numgold}={\numques}$, as follows. When $\numgold < \numques$, from a worker's point of view, the set of ${\numgold}$ questions is distributed uniformly at random in the superset of ${\numques}$ questions. However, for every set of ${\numgold}$ questions, the relations~\eqref{eq:mechanism_working_skip_ineq0},~\eqref{eq:mechanism_working_skip_ineq1},~\eqref{eq:mechanism_working_skip_ineq2} and their associated equality/strict-inequality conditions hold. The expected payment is thus (strictly) maximized when the worker answers the questions for which her confidence is greater than $T$ and skips those for which her confidence is smaller than $T$.

One can see that for every question that the worker chooses to answer, the expected payment increases with an increase in her confidence. Thus, the worker is incentivized to select the answer that she thinks is most probably correct. 

Finally, since $\kappa = \maxpay T^\numgold > 0$ and $T \in (0,1)$, the payment is always non-negative and satisfies the $\maxpay$-budget constraint.

\subsection{Uniqueness of this Mechanism}
\label{subsec:skip_unique}
While we started out with a very weak condition of no-free-lunch of that requires a minimum payment when \emph{all} attempted answers are wrong, the mechanism proposed in Algorithm~\ref{algo:incentive_skip} is significantly more strict and pays the minimum amount when \emph{any} of the attempted answers is wrong. A natural question that arises is: can we design an alternative mechanism satisfying incentive compatibility and no-free-lunch that operates somewhere in between? The following theorem answers this question in the negative.
\begin{theorem}\label{thm:skip_unique}
The mechanism of Algorithm~\ref{algo:incentive_skip} is the only incentive-compatible mechanism that satisfies the no-free-lunch condition.
\end{theorem}

Theorem~\ref{thm:skip_unique} gives a strong result despite imposing very weak requirements. To see this, recall our earlier discussion on deterring spammers, that is, incurring a low expenditure on workers who answer randomly. For instance, when the task comprises binary-choice questions, one may wish to design mechanisms which make the minimum possible payment when the responses to $50\%$ or more of the questions in the gold standard are incorrect. The no-free-lunch axiom is a much weaker requirement, and the only mechanism that can satisfy this requirement is the mechanism of Algorithm~\ref{algo:incentive_skip}.

The proof of Theorem~\ref{thm:skip_unique} is based on the following key lemma, establishing a condition that any incentive-compatible mechanism must necessarily satisfy. 
Note that this lemma does \textit{not} require the no-free-lunch condition.
\begin{lemma}
Any incentive-compatible mechanism $\payfn$ must satisfy, for every gold standard question $i \in \{1,\ldots,{\numgold}\}$ and every $(y_1,\ldots,y_{i-1},y_{i+1},\ldots,y_{\numgold} ) \in \{-1,0,1\}^{{\numgold}-1}$, 
\begin{align*}
&T f(y_1,\ldots,y_{i-1},1,y_{i+1},\ldots,y_{\numgold}) + (1-T) f(y_1,\ldots,y_{i-1},-1,y_{i+1},\ldots,y_{\numgold})\nonumber\\
&\qquad\qquad\qquad\qquad\qquad\qquad\qquad\qquad\qquad\qquad = f(y_1,\ldots,y_{i-1},0,y_{i+1},\ldots,y_{\numgold})~.
\end{align*}
\label{lem:nec_skip}
\end{lemma}
The proof of Lemma~\ref{lem:nec_skip} is provided in Appendix~\ref{app:proof_lem_nec_skip}. Using this lemma, we will now prove Theorem~\ref{thm:skip_unique}. The reader interested in further results and not the proof may feel free to jump to Subsection~\ref{sec:spamming} without any loss in continuity.

\paragraph{Proof of Theorem~\ref{thm:skip_unique}.}
The property of incentive-compatibility does not change upon any shift of the mechanism by a constant value or any scaling by a positive constant value. As a result, for the purposes of this proof, we can assume without loss of generality that $\minpay=0$. 

We will first prove that any incentive-compatible mechanism satisfying the no-free-lunch condition must make a zero payment if one or more answers in the gold standard are incorrect. The proof proceeds by induction on the number of skipped questions $S$ in the gold standard. Let us assume for now that in the ${\numgold}$ questions in the gold standard, the first question is answered incorrectly, the next $({\numgold}-1-S)$ questions are answered by the worker and have arbitrary evaluations, and the remaining $S$ questions are skipped. The proof proceeds by an induction on $S$. Suppose $S={\numgold}-1$. In this case, the only attempted question is the first question and the answer provided by the worker to this question is incorrect. The no-free-lunch condition necessitates a zero payment in this case, thus satisfying the base case of our induction hypothesis. Now we prove the hypothesis for some $S$ under the assumption of it being true when the number of questions skipped in the gold standard is $(S+1)$ or more. From Lemma~\ref{lem:nec_skip}  (with $i={\numgold}-S-1$) we have
\begin{align*}
&T f(-1,y_2,\ldots,y_{{\numgold}-S-2},1,0,\ldots,0) + (1-T) f(-1,y_2,\ldots,y_{{\numgold}-S-2},-1,0,\ldots,0)\nonumber\\
&\qquad\qquad\qquad\qquad\qquad\qquad\qquad\qquad\qquad\qquad\qquad\qquad = f(-1,y_2,\ldots,y_{{\numgold}-S-2},0,0,\ldots,0)~\nonumber\\
&\qquad\qquad\qquad\qquad\qquad\qquad\qquad\qquad\qquad\qquad\qquad\qquad =0,\end{align*}
where the final equation is a consequence of our induction hypothesis: The induction hypothesis is applicable since $f(-1,y_2,\ldots,y_{{\numgold}-S-2},0,0,\ldots,0)$ corresponds to the case when the last $(S+1)$ questions are skipped and the first question is answered incorrectly. Now, since the payment $f$ must be non-negative and since $T \in (0,1)$, it must be that 
\beq f(-1,y_2,\ldots,y_{{\numgold}-S-2},1,0,\ldots,0)=0, 
\nonumber \eeq and 
\begin{align*} 
f(-1,y_2,\ldots,y_{{\numgold}-S-2},-1,0,\ldots,0)=0.
\end{align*} This completes the proof of our induction hypothesis. Furthermore, each of the arguments above hold for any permutation of the ${\numgold}$ questions, thus proving the necessity of zero payment when any one or more answers are incorrect.

We will now prove that when no answers in the gold standard are incorrect, the payment must be of the form described in Algorithm~\ref{algo:incentive_skip}. Let $\kappa$ be the payment when all ${\numgold}$ questions in the gold standard are skipped. Let $C$ be the number questions answered correctly in the gold standard. Since there are no incorrect answers, it follows that the remaining $({\numgold}-C)$ questions are skipped. Let us assume for now that the first $C$ questions are answered correctly and the remaining $({\numgold}-C)$ questions are skipped. We repeatedly apply Lemma~\ref{lem:nec_skip}, and the fact that the payment must be zero when one or more answers are wrong, to get
\begin{align*}
f(\underbrace{1,\ldots,1}_{C-1},1,\underbrace{0,\ldots,0}_{{\numgold}-C}) &= \frac{1}{T} f(\underbrace{1,\ldots,1}_{C-1},0,\underbrace{0,\ldots,0}_{{\numgold}-C}) - \frac{1-T}{T} f(\underbrace{1,\ldots,1}_{C-1},-1,\underbrace{0,\ldots,0}_{{\numgold}-C})\\
&=\frac{1}{T} f(\underbrace{1,\ldots,1}_{C-1},0,\underbrace{0,\ldots,0}_{{\numgold}-C})\\
&\quad \vdots\\
&=\frac{1}{T^C} f(\underbrace{0,\ldots,0}_{{\numgold}})\\
&= \frac{1}{T^C} \kappa~.
\end{align*}
In order to abide by the budget, we must have the maximum payment as $\maxpay = \kappa \frac{1}{T^\numgold}$. It follows that $\kappa = \maxpay T^\numgold$. Finally, the arguments above hold for any permutation of the ${\numgold}$ questions, thus proving the uniqueness of the mechanism of Algorithm~\ref{algo:incentive_skip}.

\subsection{Optimality against Spamming Behavior}
\label{sec:spamming}
As discussed earlier, crowdsouring tasks, especially those with multiple choice questions, often encounter spammers who answer randomly without heed to the question being asked. For instance, under a binary-choice setup, a spammer will choose one of the two options uniformly at random for every question. A highly desirable objective in crowdsourcing settings is to deter spammers. To this end, one may wish to impose a condition of making the minimum possible payment when the responses to $50\%$ or more of the attempted questions in the gold standard are incorrect.
A second desirable metric could be to minimize the expenditure on a worker who simply skips all questions. While the aforementioned requirements were deterministic functions of the worker's responses, one may alternatively wish to impose requirements that depend on the distribution of the worker's answering process. For instance, a third desirable feature would be to minimize the expected payment to a worker who answers all questions uniformly at random. We now show that interestingly, our unique multiplicative payment mechanism \emph{simultaneously} satisfies all these requirements. The result is stated assuming a multiple-choice setup, but extends trivially to non-multiple-choice settings.

\begin{subtheorem}{theorem}\label{prop:skip_spammer}
\begin{theorem}[Distributional] Consider any value $A \in \{0,\ldots,\numgold\}$. Among all incentive-compatible mechanisms (that may or may not satisfy no-free-lunch), Algorithm~\ref{algo:incentive_skip} strictly minimizes the expenditure on a worker who skips some $A$ of the questions in the the gold standard, and chooses answers to the remaining $(\numgold-A)$ questions uniformly at random.
\end{theorem}
\begin{theorem}[Deterministic] Consider any value $B \in (0,1]$. Among all incentive-compatible mechanisms (that may or may not satisfy no-free-lunch), Algorithm~\ref{algo:incentive_skip} strictly minimizes the expenditure on a worker who gives incorrect answers to a fraction $B$ or more of the questions attempted in the gold standard.
\end{theorem}
\end{subtheorem}
We see from this result that the multiplicative payment mechanism of Algorithm~\ref{algo:incentive_skip} thus possesses very useful properties geared to deter spammers, while ensuring that a good worker will be paid a high enough amount. 
To illustrate this point, let us compare the mechanism of Algorithm~\ref{algo:incentive_skip} with the popular additive class of payment mechanisms. 
\begin{example} Consider the popular class of ``additive'' mechanisms, where the payments to a worker are added across the gold standard questions. This additive payment mechanism offers a reward of $\frac{\maxpay}{\numgold}$ for every correct answer in the gold standard, $\frac{\maxpay T}{\numgold}$ for every question skipped, and $0$ for every incorrect answer. Importantly, the final payment to the worker is the \emph{sum} of the rewards across the $\numgold$ gold standard questions. One can verify that this additive mechanism is incentive compatible. One can also see that that as guaranteed by our theory, this additive payment mechanism does not satisfy the no-free-lunch axiom. 

Suppose each question involves choosing from two options. Let us compute the expenditure that these two mechanisms make under a spamming behavior of choosing the answer randomly to each question. Given the $50\%$ likelihood of each question being correct, on can compute that the additive mechanism makes a payment of $\frac{\maxpay}{2}$ in expectation. On the other hand, our mechanism pays an expected amount of only $\maxpay 2^{-\numgold}$. The payment to spammers thus reduces exponentially with the number of gold standard questions under our mechanism, whereas it does not reduce at all in the additive mechanism.

Now, consider a different means of exploiting the mechanism(s) where the worker simply skips all questions. To this end, observe that if a worker skips all the questions then the additive payment mechanism will incur an expenditure of $\maxpay T$. On the other hand, the proposed payment mechanism of Algorithm~\ref{algo:incentive_skip} pays an exponentially smaller amount of $\maxpay T^{\numgold}$ (recall that $T<1$).
\end{example}

We prove Theorem~\ref{prop:skip_spammer} in the rest of this subsection. The reader may feel free to jump directly to Section~\ref{sec:confidence} without any loss in continuity.

\paragraph{Proof of Theorem~\ref{prop:skip_spammer}.} 
The property of incentive-compatibility does not change upon any shift of the mechanism by a constant value or any scaling by a positive constant value. As a result, for the purposes of this proof, we can assume without loss of generality that $\minpay=0$. 

{\bf Part A (Distributional).} Let $m$ denote the number of options in each question. One can verify that under the mechanism of Algorithm~\ref{algo:incentive_skip}, a worker who skips $A$ questions and answers the rest uniformly at random will get a payment of $\frac{\maxpay T^A}{m^{\numgold-A}}$ in expectation. This expression arises due to the fact that Algorithm~\ref{algo:incentive_skip} makes a zero payment if any of the attempted answers are incorrect, and a payment of $\maxpay T^A$ if the worker skips $A$ questions and answers the rest correctly. Under uniformly random answers, the probability of the latter event is $\frac{1}{m^{\numgold-A}}$.

Now consider any other mechanism, and denote it as $f'$. Let us suppose without loss of generality that the worker attempts the first $(G-A)$ questions. Since the payment must be non-negative, a repeated application of Lemma~\ref{lem:nec_skip} gives
\begin{align}
f'(\underbrace{1,\ldots,1}_{G-A},0,\ldots,0) 
&\geq T f'(\underbrace{1,\ldots,1}_{G-A + 1},0,\ldots,0) \label{eq:minspam_dist0} \\
& \quad \vdots \nonumber \\
&\geq T^A f'(1,\ldots,1) \nonumber \\
& = T^A \maxpay, \label{eq:minspam_dist2}
\end{align}
where~\eqref{eq:minspam_dist2} is a result of the $\maxpay$-budget constraint. Since there is a $\frac{1}{m^{\numgold-A}}$ chance of the $(G-A)$ attempted answers being correct, the expected payment under any other mechanism $f'$ must be at least $\frac{\maxpay T^A}{m^{\numgold-A}}$. 

We will now show that if any mechanism $f'$ that makes an expected payment of $\frac{\maxpay T^A}{m^{\numgold-A}}$ to such a spammer, then the mechanism must be identical to Algorithm~\ref{algo:incentive_skip}. We split the proof of this part into two cases, depending on the value of the parameter $A$.

Case I ($A<G)$: In order to make an expected payment of $\frac{\maxpay T^A}{m^{\numgold-A}}$, the mechanism must achieve the bound~\eqref{eq:minspam_dist2} with equality, and furthermore, the mechanism must have zero payment if any of the $(G-A)$ attempted questions are answered incorrectly. In other words, the mechanism $f'$ under consideration must satisfy
\begin{align*}
f'(y_1,\ldots,y_{G-A},0,\ldots,0) = 0 \qquad \forall (y_1,\ldots,y_{G-A}) \in \{-1,1\}^{G-A} \backslash \{1\}^{G-A}.
\end{align*}
A repeated application of Lemma~\ref{lem:nec_skip} then implies
\begin{align}
f'(0,0,\ldots,-1) = 0.
\label{eq:minspam_dist3}
\end{align}
Note that so far we considered the case when the worker attempts the first $(G-A)$ questions. The arguments above hold for any choice of the $(G-A)$ attempted questions, and consequently the results shown so far in this proof hold for all permutations of the arguments to $f'$. In particular, the mechanism $f'$ must make a zero payment when any $(G-1)$ questions in the gold standard are skipped and the remaining question is answered incorrectly. Another repeated application of Lemma~\ref{lem:nec_skip} to this result gives
\begin{align*}
f'(y_1,\ldots,y_G) = 0 \qquad \forall (y_1,\ldots,y_G) \in \{0,-1\}^\numgold \backslash \{0\}^\numgold.
\end{align*}
This condition is precisely the no-free-lunch axiom, and in Theorem~\ref{thm:skip_unique} we had shown that Algorithm~\ref{algo:incentive_skip} is the only incentive-compatible mechanism that satisfies this axiom. It follows that our mechanism, Algorithm~\ref{algo:incentive_skip} strictly minimizes the expected payment in the setting under consideration.

Case II ($A = G$): In order to achieve the bound~\eqref{eq:minspam_dist2} with equality, the mechanism $f'$ must also achieve the bound~\eqref{eq:minspam_dist0} with equality. Noting that we have $A=G$ in this case, it follows that the mechanism $f'$ must satisfy
\begin{align*}
f'(-1,0,\ldots,0) = 0.
\end{align*}
This condition is identical to~\eqref{eq:minspam_dist3} established for Case I earlier, and the rest of the argument now proceeds in a manner identical to the subsequent arguments in Case I.

{\bf Part B (Deterministic).} Given our result of Theorem~\ref{thm:skip_unique}, the proof for the deterministic part is straightforward. Algorithm~\ref{algo:incentive_skip} makes a payment of zero when one or more of the answers to questions in the gold standard are incorrect. Consequently, for every value of parameter $B \in (0,1]$, Algorithm~\ref{algo:incentive_skip} makes a zero payment when a fraction $B$ or more of the attempted answers are incorrect. Any other mechanism doing so must satisfy the no-free-lunch axiom. In Theorem~\ref{thm:skip_unique} we had shown that Algorithm~\ref{algo:incentive_skip} is the only incentive-compatible mechanism that satisfies this axiom. It follows that our mechanism, Algorithm~\ref{algo:incentive_skip}, strictly minimizes the payment in the event under consideration.

\section{
Confidence-based Setting}\label{sec:confidence}
In this section, we will discuss incentive mechanisms when the worker is asked to select from more than one confidence-level for every question (Figure~\ref{fig:intro}c). In particular, for some $L \geq 1$, the worker is asked to indicate a confidence-level in the range $\{0,\ldots,L\}$ for every answer. Level $0$ is the `skip' level, and level $L$ denotes the highest confidence. Note that we do not solicit an answer if the worker indicates a confidence-level of $0$ (skip), but the worker must provide an answer if she indicates a confidence-level of $1$ or higher. This makes the case of having only a `skip' as considered in Section~\ref{sec:skip} a special case of this setting, and corresponds to $L=1$.

We generalize the requirement of no-free-lunch to the confidence-based setting as follows.
\begin{axiom}[\bf Generalized-no-free-lunch axiom] If {all} the answers attempted by the worker in the gold standard are selected as the highest confidence-level (level $L$), and {all} of them turn out to be wrong, then the payment is $\minpay$. More formally, we require the mechanism $\payfn$ to satisfy $\payfn(\ans{1},\ldots,\ans{\numgold})= \minpay$ for every evaluation $(\ans{1},\ldots,\ans{\numgold})$ that satisfies $0 < \sum_{i=1}^{\numgold} \mathbf{1}\{\ans{i} \neq 0\} = \sum_{i=1}^{\numgold} \mathbf{1}\{\ans{i} = -L\}$.
\end{axiom}

In the confidence-based setting, we require specification of a set of thresholds $\{ S_\elll,T_\elll  \}_{\elll=1}^{L}$ that determine the confidence-levels that the workers should indicate. In particular, we will require specification of two reference points for each confidence level, and this specification generalizes the skip-based setting.
\begin{itemize}
\item The first set of thresholds specifies a comparison of any confidence level with the skipping option as a fixed reference. To this end, recall that in the skip-based setting, the threshold $T$ specified when the worker should skip a question and when she should attempt to answer. This is generalized to the confidence-based setting where for every level $\elll \in [L]$, a fixed threshold $S_\elll$ specifies the `strength' of confidence-level $\elll$: If restricted to only the two options of skipping or selecting confidence-level $\elll$ for any question, the worker should be incentivized to select confidence-level $\elll$ if her confidence is higher than $S_\elll$ and skip if her confidence is lower than $S_\elll$.
\item The second set of thresholds specifies a comparison of any confidence level with its neighbors. If a worker decides to not skip a question, she must choose one of multiple confidence-levels. A set $\{ T_\elll \}_{\elll=1}^{L}$ of thresholds specify the boundaries between different confidence-levels. In particular, when the confidence of the worker for a question lies in $(T_{\elll-1},T_{\elll+1})$, then the worker must be incentivized to indicate confidence-level $(\elll-1)$ if her confidence is lower than $T_\elll$ and to indicate confidence-level $\elll$ if her confidence is higher than $T_\elll$. 
This includes selecting level $L$ if her confidence is higher than $T_L$ and selecting level $0$ if her confidence is lower than $T_1$.~
\end{itemize}
We will call a payment mechanism as incentive-compatible if it satisfies the two requirements listed above, and also incentivizes the worker to select the answer that she believes is most likely to be correct for every question for which her confidence is higher than $T_1$.

The problem setting inherently necessitates certain restrictions in the choice of the thresholds. Since we require the worker to choose a higher level when her confidence is higher, the thresholds must necessarily be monotonic and satisfy $0 < S_1 < S_2 < \cdots <S_L<1$ and $0 < T_1 < T_2 < \cdots <T_L < 1$. Also observe that the definitions of $S_1$ and $T_1$ coincide, and hence $S_1 = T_1$. Additionally, we can show (Proposition~\ref{prop:threshold_T_gt_S} in Appendix~\ref{app:SmorethanT}) that for incentive-compatible mechanisms to exist, it must be that $T_\elll > S_\elll~\forall~\elll\in\{2,\ldots,L\}$. As a result, the thresholds must also satisfy $T_1=S_1,~T_2 > S_2,\ldots,T_L > S_L$. These thresholds may be chosen based on various factors of the problem at hand, for example, on the post-processing algorithms, any statistics on the distribution of worker abilities, budget constraints, etc. In this paper, we will assume that these values are given to us.

\subsection{Payment Mechanism}\label{sec:confidence_pay}
The proposed payment mechanism is described in Algorithm~\ref{algo:incentive_confidence}.

\begin{algorithm}[H]
\begin{itemize}[leftmargin = -5pt]
\item Inputs:
\begin{itemize}[leftmargin = 12pt]
\item Thresholds $S_1,\ldots,S_L$ and $T_1,\ldots,T_L$
\item Budget parameters $\maxpay$ and $\minpay$
\item Evaluations $(\ans{1},\ldots,\ans{\numgold}) \in \{-L,\ldots,+L\}^\numgold$ of the worker's answers to the $\numgold$ gold standard questions
\end{itemize}
\item Set $\alpha_{-L},\ldots,\alpha_L$ as
\begin{itemize}[leftmargin = 12pt]
\item $\alpha_L = \frac{1}{S_L}$, $\alpha_{-L}=0$
\item For $\elll \in \{L-1,\ldots,1\}$,
\[ \alpha_{\elll} = \frac{(1-S_\elll)T_{\elll+1}\alpha_{\elll+1}+(1-S_\elll)(1-T_{\elll+1})\alpha_{-(\elll+1)}-(1-T_{\elll+1})}{T_{\elll+1}-S_\elll}
\quad\textrm{and}\quad
\alpha_{-\elll} = \frac{1-S_\elll \alpha_\elll}{1-S_\elll}
\]
\item[{\scriptsize$\blacktriangleright$}] $\alpha_0=1$
\end{itemize}
\item The payment is
\begin{equation*}
f(\ans{1},\ldots,\ans{\numgold}) = \kappa \prod_{i=1}^{\numgold} \alpha_{\ans{i}} + \minpay
\end{equation*}
where $\kappa = (\maxpay - \minpay) \left(\frac{1}{\alpha_L}\right)^\numgold$.
\end{itemize}
\caption{Incentive mechanism for the confidence-based setting}
\label{algo:incentive_confidence}
\end{algorithm}

The following theorem shows that this mechanism indeed incentivizes a worker to select answers and confidence-levels as desired. 
\begin{theorem}
The mechanism of Algorithm~\ref{algo:incentive_confidence} is incentive-compatible and satisfies the generalized-no-free-lunch condition.
\label{thm:mechanism_working_confidence}
\end{theorem}
The proof of Theorem~\ref{thm:mechanism_working_confidence} follows in a manner similar to that of the proof of Theorem~\ref{thm:mechanism_working_skip}, and is provided in Appendix~\ref{app:proof_incentive_confidence}.

\begin{remark}\label{cor:mechanism_working_confidence}
The mechanism of Algorithm~\ref{algo:incentive_confidence} also ensures a condition stronger than the `boundary-based' definition of the thresholds $\{T_\elll\}_{\elll \in [L]}$ given earlier. Under this mechanism, for every $\elll \in [L-1]$ the worker is incentivized to select confidence-level $\elll$ (over all else) whenever her confidence lies in the interval $(T_\elll,T_{\elll+1})$, select confidence-level $0$ (over all else) whenever her confidence is lower than $T_1$ and select confidence-level $L$ (over all else) whenever her confidence is higher than $T_L$. 
\end{remark}

\subsection{Uniqueness of this Mechanism}
We prove that the mechanism of Algorithm~\ref{algo:incentive_confidence} is unique, that is, no other incentive-compatible mechanism can satisfy the generalized-no-free-lunch condition. 
\begin{theorem}\label{thm:unique_confidence}
The payment mechanism of Algorithm~\ref{algo:incentive_confidence} is the only incentive-compatible mechanism that satisfies the generalized-no-free-lunch condition.
\end{theorem}

The proof of Theorem~\ref{thm:unique_confidence} is provided in Appendix~\ref{app:proof_unique_confidence}. The proof is conceptually similar to that of Theorem~\ref{thm:unique_confidence} but involves resolving several additional complexities that arise due to elicitation from multiple confidence levels. 

\section{A Stronger No-free-lunch Condition: Impossibility Results}\label{sec:strong_skip}
Recall that the no-free-lunch axiom under the skip-based mechanism of Section~\ref{sec:skip} requires the payment to be the minimum possible if all attempted answers in the gold standard are incorrect. However, a worker who skips all the questions may still receive a payment. The generalization under the confidence-based mechanism of Section~\ref{sec:confidence} requires the payment to be the minimum possible if all attempted answers in the gold standard were selected with the highest confidence-level and were incorrect. However, a worker who marked all questions with a lower confidence level may be paid even if her answers to all the questions in the gold standard turn out to be incorrect. One may thus wish to impose a stronger requirement instead, where the minimum payment is made to workers who make no useful contribution. This is the primary focus of this section.

Consider the skip-based setting. Define the following axiom which is slightly stronger than the no-free-lunch axiom defined previously.

\textit{Strong-no-free-lunch}: If none of the answers in the gold standard are correct, then the payment is $\minpay$. More formally, strong-no-free-lunch imposes the condition $\payfn(\ans{1},\ldots,\ans{\numgold}) = \minpay$ for every evaluation $(\ans{1},\ldots,\ans{\numgold})$ that satisfies $\sum_{i=1}^{\numgold} \mathbf{1}\{\ans{i}>0\} = 0$.

The strong-no-free-lunch axiom is only slightly stronger than the no-free-lunch axiom proposed in Section~\ref{sec:skip} for the skip-based setting. The strong-no-free-lunch condition can equivalently be written as imposing requiring the payment to be the minimum possible for every evaluation that satisfies  $\sum_{i=1}^{\numgold} \mathbf{1}\{\ans{i} \neq 0\} = \sum_{i=1}^{\numgold} \mathbf{1}\{\ans{i} = -1\}$. From this interpretation, one can see that to the set of events necessitating the minimum payment under the no-free-lunch axiom, the strong-no-free-lunch axiom adds only one extra event, the event of the worker skipping all questions. Unfortunately, it turns out that this minimal strengthening of the requirements is associated to impossibility results.

In this section we show that no mechanism satisfying the strong-no-free-lunch axiom can be incentive compatible in general. The only exception is the case when (a) all questions are in the gold standard (${\numgold}={\numques}$), and (b) it is guaranteed that the worker has a confidence greater than $T$ for at least one of the $\numques$ questions. These conditions are, however, impractical for the crowdsourcing setup under consideration in this paper. We will first prove the impossibility results under the strong-no-free-lunch axiom. For the sake of completeness (and also to satisfy mathematical curiosity), we will then provide a (unique) mechanism that is incentive-compatible and satisfies the strong-no-free-lunch axiom for the skip-based setting under the two conditions listed above. The proofs of each of the claims made in this section are provided in Appendix~\ref{app:proof_strong_skip}.

Let us continue to discuss the skip-based setting. In this section, we will call any worker whose confidences for all of the $\numques$ questions is lower than $T$ as an \textit{unknowledgeable worker}, and call the worker a \textit{knowledgeable worker} otherwise.

\begin{proposition}\label{prop:strong_unknowledgeable}
No payment mechanism satisfying the strong-no-free-lunch condition can incentivize an unknowledgeable worker to skip all questions. As a result, no mechanism satisfying the strong-no-free-lunch axiom can be incentive-compatible. 
\end{proposition}
The proof of this proposition, and that of all other theoretical claims made in this section, are presented in Appendix~\ref{app:proof_strong_skip}. 

The impossibility result of Proposition~\ref{prop:strong_unknowledgeable} relies on trying to incentivize an unknowledgeable worker to act as desired. Since no mechanism can be incentive compatible for unknowledgeable workers, we will now consider only workers who are knowledgeable. The following proposition shows that the strong-no-free-lunch condition is too strong even for this relaxed setting.
\begin{proposition}\label{prop:strong_knowledgeable}
When ${\numgold}<{\numques}$, there exists no mechanism that is incentive-compatible for knowledgeable workers and satisfies the strong-no-free-lunch condition.
\end{proposition}
Given this impossibility result for ${\numgold}<{\numques}$, we are left with ${\numgold}={\numques}$ which means that the true answers to all the questions are known apriori. This condition is clearly not applicable to a crowdsourcing setup; nevertheless, it is mathematically interesting and may be applicable to other scenarios such as testing and elicitation of beliefs about future events. 

Proposition~\ref{prop:mechanism_working_skip_strong} below presents a mechanism for this case and proves its uniqueness. We previously saw that an unknowledgeable worker cannot be incentivized to skip all the questions (even when $\numgold=\numques$). Thus, in our payment mechanism, we do the next best thing: Incentivize the unknowledgeable worker to answer only one question, that which she is most confident about, while incentivizing the knowledgeable worker to answer questions for which her confidence is greater than $T$ and skip those for which her confidence is smaller than $T$. 

\begin{proposition}
Let $C$ be the number of correct answers and $W$ be the number of wrong answers (in the gold standard). Let the payment be $\minpay$ if $W>0$ or $C=0$, and be $(\maxpay - \minpay) T^{\numgold-C} + \minpay$ otherwise. Under this mechanism, when ${\numgold}={\numques}$, an unknowledgeable worker is incentivized to answer only one question, that for which her confidence is the maximum, and a knowledgeable worker is incentivized to answer the questions for which her confidence is greater than $T$ and skip those for which her confidence is smaller than $T$. Furthermore, when ${\numgold}={\numques}$, this mechanism is the one and only mechanism that obeys the strong-no-free-lunch condition and is incentive-compatible for knowledgeable workers.
\label{prop:mechanism_working_skip_strong}
\end{proposition}

The following proposition shows that the strong-no-free-lunch condition leads to negative results in the confidence-based setting ($L>1$) as well. The strong-no-free-lunch condition is still defined as in the beginning of Section~\ref{sec:strong_skip}, i.e., the payment is zero if none of the answers are correct.
\begin{proposition}\label{prop:strong_confidence}
When $L>1$, for any values of $\numques$ and $\numgold~(\leq \numques)$, it is impossible for any mechanism to satisfy the strong-no-free-lunch condition and be incentive-compatible even when the worker is knowledgeable.
\end{proposition}

\section{Simulations and Experiments}\label{sec:sim_exp}

In this section, we present synthetic simulations and real-world experiments to evaluate the effects of our setting and our mechanism on the final label quality.

\subsection{Synthetic Simulations}\label{sec:simulations}
We employ synthetic simulations to understand the effects of various distributions of the confidences and labeling errors. We consider binary-choice questions in this set of simulations. Whenever a worker answers a question, her confidence for the correct answer is drawn from a distribution $\mathcal{P}$ independent of all else. We investigate the effects of the following five choices of the distribution $\mathcal{P}$:
\begin{itemize}
\item The uniform distribution on the support $[0.5,1]$.
\item A triangular distribution with lower end-point $0.2$, upper end-point $1$ and a mode of $0.6$.
\item A beta distribution with parameter values $\alpha = 5$ and $\beta = 1$.
\item The hammer-spammer distribution~\citepn{karger2011iterative}: uniform on the discrete set $\{0.5,1\}$.
\item A truncated Gaussian distribution: a truncation of $\mathcal{N}(0.75, 0.5)$ to the interval $[0,1]$.
\end{itemize}

We compare (a) the setting where workers attempt every question, with (b) the setting where workers skip questions for which their confidence is below a certain threshold $T$. In this set of simulations, we set $T = 0.75$. In either setting, we aggregate the labels obtained from the workers for each question via a majority vote on the two classes. Ties are broken by choosing one of the two options uniformly at random.

\begin{figure}
\centering
\includegraphics[width=.95\textwidth]{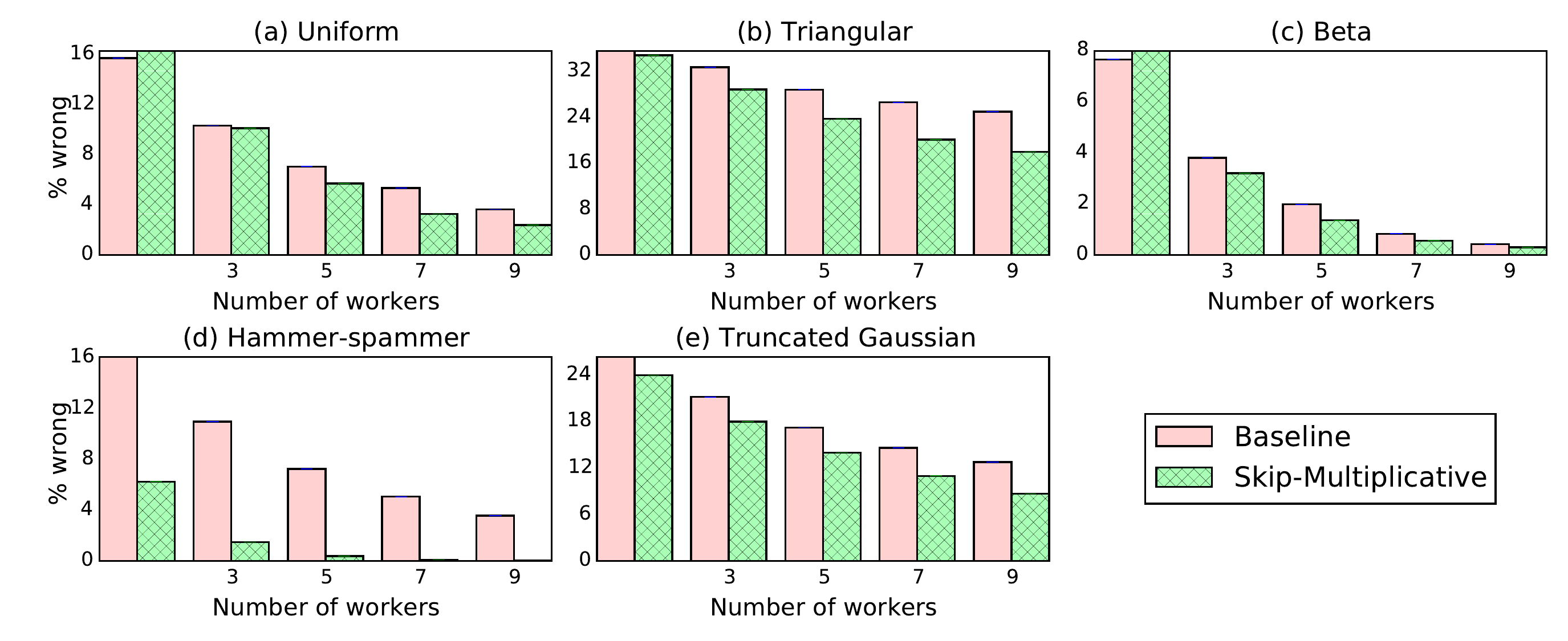}
\caption{Error under different interfaces for synthetic simulations of five distributions of the workers' error probabilities.}
\label{fig:simulations}
\end{figure}

Figure~\ref{fig:simulations} depicts the results from these simulations. Each bar represents the fraction of questions that are labeled incorrectly, and is an average across $50,\!000$ trials. (The standard error of the mean is too small to be visible.) We see that the skip-based setting consistently outperforms the conventional setting, and the gains obtained are moderate to high depending on the underlying distribution of the workers' errors. In particular, the gains are quite striking under the hammer-spammer model: this result is not surprising since the mechanism (ideally) screens the spammers out and leaves only the hammers who answer perfectly.

\begin{figure}
\centering
\includegraphics[width=.95\textwidth]{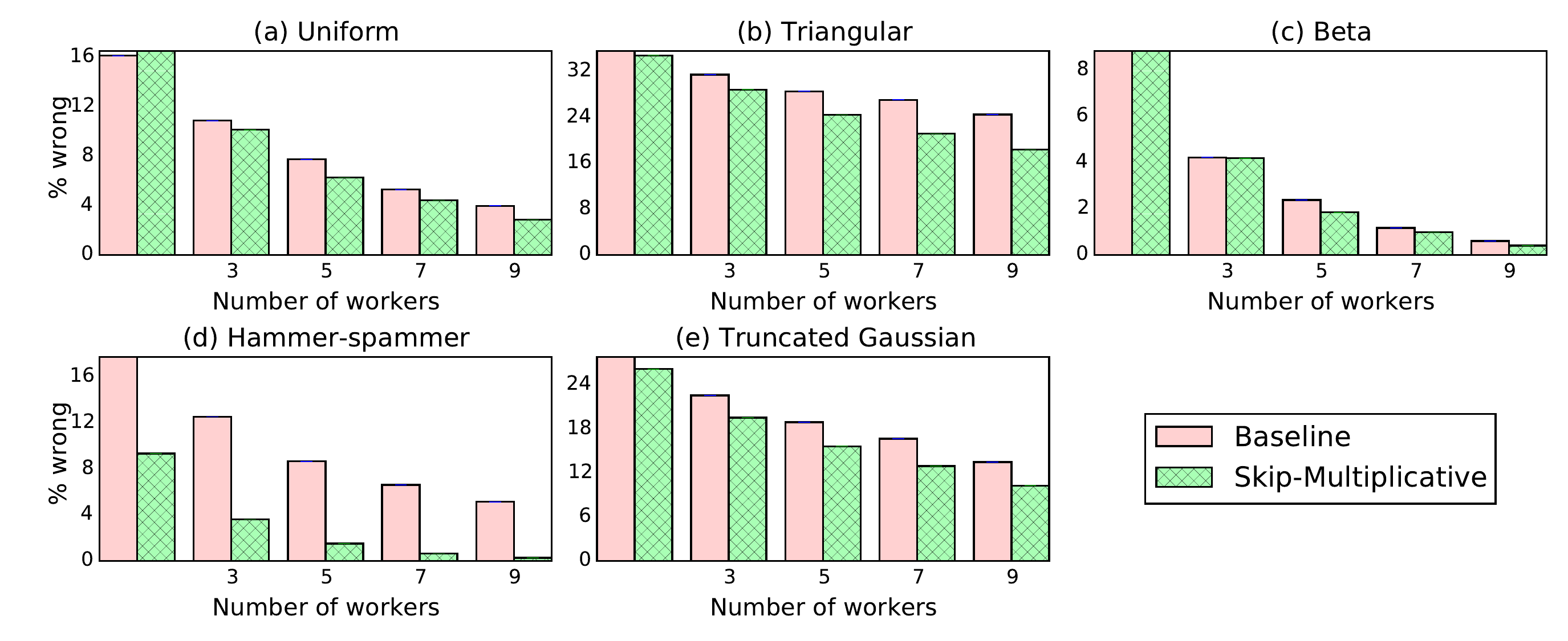}
\caption{Errors under a model that is a perturbation of the first experiment, where the worker's confidence is a noisy version of the true error probability and where ties are considered different from random decisions.}
\label{fig:simulations_perturb}
\end{figure}

The setup of the simulations described above assumes that the workers confidences equal the true error probabilities. In practice, however, the workers may have incorrect beliefs. The setup also assumes that ties are broken randomly; however in practice, ties may be broken in a more systematic manner by eliciting additional labels for only these hard questions. We now present a second set of simulations that mitigates these biases. In particular, when a worker has a confidence of $p$, the actual probability of error is assumed to be drawn from a Gaussian distribution with mean $p$ and standard deviation $0.1$, truncated to $[0,1]$. In addition, when evaluating the performance of the majority voting procedure, we consider a tie as having an error of $0.4$. Figure~\ref{fig:simulations_perturb} depicts the results of these simulations. We observe that the results from these simulations are very similar to those obtained in the earlier simulation setup of Figure~\ref{fig:simulations}.

\subsection{Experiments on Amazon Mechanical Turk}\label{sec:experiments}

We conducted preliminary experiments on the Amazon Mechanical Turk commercial crowdsourcing platform (\url{mturk.com}) to evaluate our proposed scheme in real-world scenarios. The complete data, including the interface presented to the workers in each of the tasks, the results obtained from the workers, and the ground truth solutions, are available on the website of the first author.

\subsubsection{Goal} 
Before delving into details, we first note certain caveats relating to such a study of mechanism design on crowdsourcing platforms. When a worker encounters a mechanism for only a small amount of time (a handful of tasks in typical research experiments) and for a small amount of money (at most a few dollars in typical crowdsourcing tasks), we cannot expect the worker to completely understand the mechanism and act precisely as required. For instance, we wouldn't expect our experimental results to change significantly even upon moderate modifications in the promised amounts, and furthermore, we do expect the outcomes to be noisy. Incentive compatibility kicks in when the worker encounters a mechanism across a longer term, for example, when a proposed mechanism is adopted as a standard for a platform, or when higher amounts are involved. This is when we would expect workers or others (e.g., bloggers or researchers) to design strategies that can game the mechanism. The theoretical guarantee of incentive compatibility then prevents such gaming in the long run.

We thus regard these experiments as preliminary. Our intentions towards this experimental exercise were (a) to evaluate the potential of our algorithms to work in practice, (b) to investigate the effect of the proposed algorithms on the net error in the collected labelled data, and (c) to identify if there is any major issue of dissatisfaction among the workers.

\subsubsection{Experimental setup} 
We conducted our experiments on the ``Amazon Mechanical Turk'' commercial crowdsourcing platform (\url{mturk.com}). On this platform, individuals or businesses (called `requesters') can post tasks, and any individual (called a `worker') may complete the task over the Internet in exchange for a pre-specified payment. The payment may comprise of two parts: a fixed component which is identical for all workers performing that task, and a `bonus' which may be different for different workers and is paid at the discretion of the requester.

We designed nine experiments (tasks) ranging from image annotation to text and speech recognition. The individual experiments are described in more detail in Appendix~\ref{app:experiments}. All experiments involved objective questions, and the responses elicited were multiple choice in five of the experiments and freeform text in the rest. For each experiment, we tested three settings: (i) the baseline conventional setting (Figure~\ref{fig:intro}a) with a mechanism of paying a fixed amount per correct answer, (ii) our skip-based setting (Figure~\ref{fig:intro}b) with our multiplicative mechanism, and (iii) our confidence-based setting (Figure~\ref{fig:intro}c) with our confidence-based mechanism. For each mechanism in each experiment, we specified the requirement of 35 workers independently performing the task. This amounts to a total of 945 worker-tasks (315 worker-tasks for each mechanism). We also set the following constraints for a worker to attempt our tasks: the worker must have completed at least 100 tasks previously, and must have a history of having at least 95$\%$ of her prior work approved by the respective requesters. In each experiment, we offered a certain small fixed payment (in order to attract the workers in the first place) and executed the variable part of our mechanisms via a bonus payment.

\subsubsection{Results: Raw data}
\begin{figure*}
\centering
\includegraphics[width=.9\textwidth]{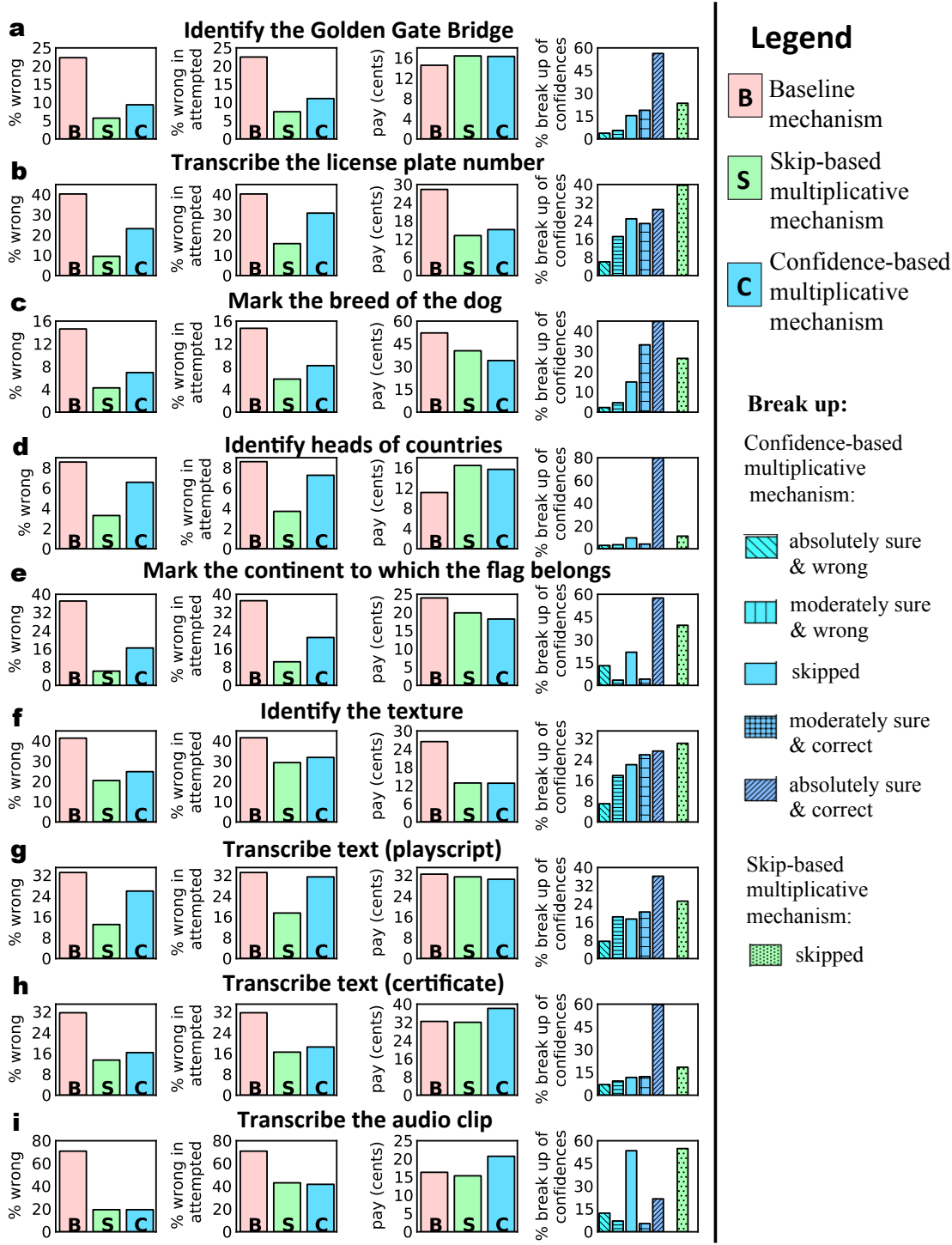}
\caption{The error-rates in the raw data and payments in the nine experiments. Each individual bar in the plots corresponds to one mechanism in one experiment and is generated from 35 distinct workers (this totals to 945 worker-tasks).}
\label{fig:plots_supplement}
\end{figure*}

Figure~\ref{fig:plots_supplement} plots, for the baseline, skip-based and confidence-based mechanisms for all nine experiments, the (i) fraction of questions that were answered incorrectly, (ii) fraction of questions that were answered incorrectly among those that were attempted, (iii) the average payment to a worker (in cents), and (iv) break up of the answers in terms of the fraction of answers in each confidence level. The payment for various tasks plotted in Figure~\ref{fig:plots_supplement} is computed as the average of the payments across $100$ (random) selections of the gold standard questions, in order to prevent any distortion of the results due to the randomness in the choice of the gold standard questions. 

The figure shows that the amount of errors among the attempted questions is much lower in the skip and the confidence-based settings than the baseline setting. Also observe that in the confidence-based setting, as expected, the answers selected with higher confidence-levels are more correct. The total expenditure under each of these settings is similar, with the skip and the confidence-based settings faring better in most cases. We also elicited feedback from the workers, in which we received several positive comments (and no negative comments). Examples of comments that we received: ``I was wondering if it would possible to increase the maximum number of HITs I may complete for you.  As I said before, they were fun to complete.  I think I did a good job completing them, and it would be great to complete some more for you.''; ``I am eagerly waiting for your bonus.''; ``Enjoyable. Thanks.''

\subsubsection{Results: Aggregated data}
We saw in the previous section that under the skip-based setting, the amount of error among the attempted questions was significantly lower than the amount of error in the baseline setting. However, the skip-based setting was also associated, by design, to lesser amount of data by virtue of questions being skipped by the workers. A natural question that arises is how the baseline and the skip-based mechanisms will compare in terms of the final data quality, i.e., the amount of error once data from multiple workers is aggregated.

To this end, we considered the five experiments that consisted of multiple-choice questions. We let a parameter {\tt num\_workers} take values in $\{3,\,5,\,7,\,9,\,11\}$. For each of the five experiments and for each of the five values of {\tt num\_workers}, we perform the following actions $1,\!000$ times: for each question, we choose {\tt num\_workers} workers and perform a majority vote on their responses. If the correct answer for that question does not lie in the set of options given by the majority, we consider it as an accuracy of zero. Otherwise, if there are $m$ options tied in the majority vote, and the correct answer is one of these $m$, then we consider it as an accuracy of $\frac{100}{m} \%$  (hence, $100\%$ if the correct answer is the only answer picked by the majority vote). We average the accuracy across all questions and across all iterations.

We choose majority voting as the means of aggregation since (a) it is the simplest and still most popular aggregation method, and (b) to enable an apples-to-apples comparison design  since while more advanced aggregation algorithms have been developed for the baseline setting without the skip~\citepn{raykar2010learning,zhou2012learning, wauthier2011bayesian, chen2013pairwise,kamar2012combining, dawid1979maximum, karger2011iterative,liu2012variational, vempaty2014reliable, zhang2014spectral, ipeirotis2014repeated}, design of analogous algorithms for the new skip-based setting hasn't been explored yet.

\begin{figure*}
\centering
\includegraphics[width=\textwidth]{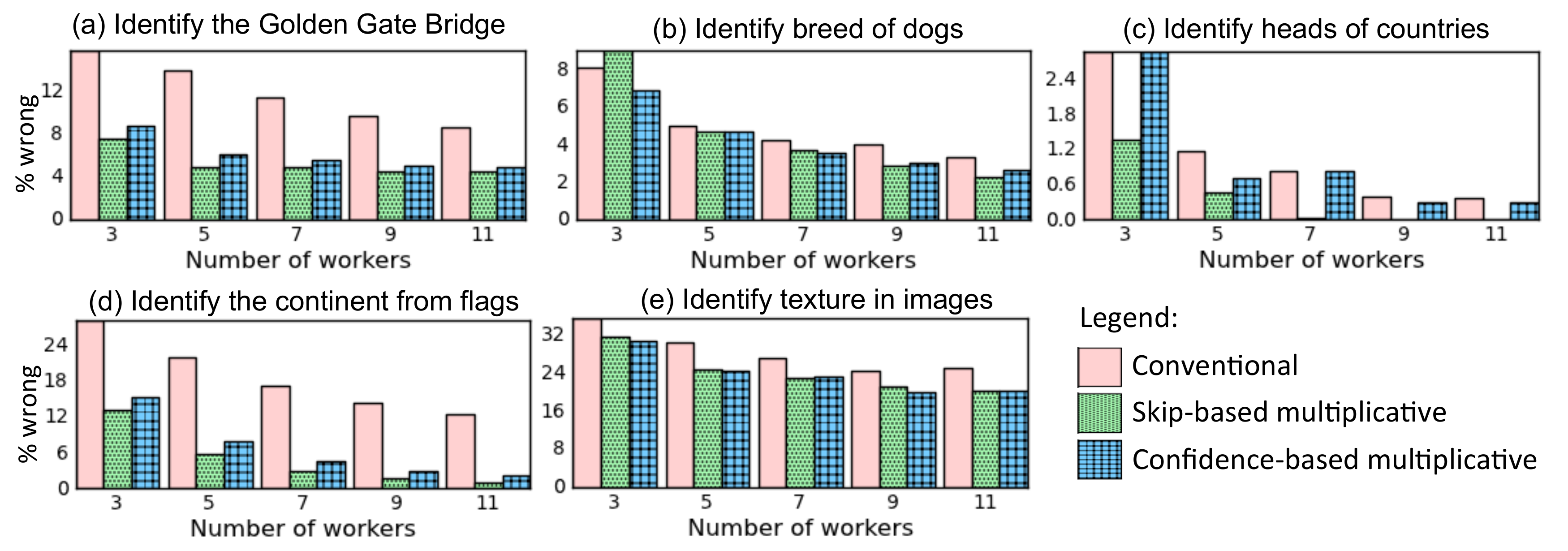}
\caption{Error-rates in the aggregated data in the five experiments involving multiple-choice tasks.}
\label{fig:plots_aggregate}
\end{figure*}

The results are presented in Figure~\ref{fig:plots_aggregate}. We see that in most cases, our skip-based mechanism induces a lower labelling error at the aggregate level than the baseline. Furthermore, in many of the instances, the reduction is two-fold or higher.

All in all, in the experiments, we observe a substantial reduction in the error-rates while expending the same or lower amounts and receiving no negative comments from the workers, suggesting that these mechanisms can work; the fundamental theory underlying the mechanisms ensures that the system cannot be gamed in the long run. Our proposed settings and mechanisms thus have the potential to provide much higher quality labeled data as input to machine learning algorithms.

\section{Discussion and Conclusions}\label{sec:conclusions}

In this concluding section, we first discuss the modelling assumptions that we made in this paper, followed by a discussion on future work and concluding remarks.

\subsection{Modelling Assumptions}
When forming the model for our problem, as in any other field of theoretical research, we had to make certain assumptions and choices. In what follows, we discuss the reasons for the modelling choices we made.

\begin{itemize}[leftmargin=12pt]
\item \emph{Use of gold standard questions.}
We assume the existence of gold standard questions in the task, i.e., a subset of questions to which the answers are known to the system designer. The existence of gold standard is commonplace in crowdsourcing platforms~\citepn{le2010ensuring,chen2011opportunities}.



\item \emph{Workers aiming to maximize their expected payments:}
We assume that the workers aim to maximize their expected payments. In many other problems in game theory, one often makes the assumption that people are ``risk-averse'', and aim to maximize the expected value of some ``utility function'' of their payments. While we extend our results to general utility functions in Appendix~\ref{sec:utility} in order to accommodate such requirements, we also think that the assumption of workers maximizing their expected payments is a perfectly reasonable assumption for the crowdsourcing settings considered here. The reason is that each such task lasts for a handful of minutes and is worth a few tens of cents. Workers typically perform tens to hundreds of tasks per day, and consequently their empirical hourly wages very quickly converge to their expectation.

\item \emph{Workers knowing their confidences:}
We understand that in practice the workers will have noisy or granular estimates of their own beliefs. The mathematical assumption of workers knowing their precise confidences is an idealization intended for mathematical tractability. This is one of the reasons why we only elicit a quantized value of the workers' beliefs (in terms of skipping or choosing one of a finite number of confidence levels), and not try to ask for a precise value.

\item \emph{Eliciting a quantized version of the beliefs:}
We do not directly attempt to elicit the values of the beliefs of the workers, but instead ask them to indicate only a quantization (e.g., ``I'm not sure'' or ``moderately confident'', etc.). We prefer this quantization to direct assessment to real-valued probability, motivated by the extensive literature in psychology on the coarseness of human perception and processing (e.g.,~\citepn{miller1956magical,shiffrin1994seven,jones2013optimal,shah2015estimation}) establishing that humans are more comfortable at providing quantized responses. This notion is verified by experiments on Amazon Mechanical Turk in~\citetn{shah2015estimation} where it is observed that people are more consistent when giving ordinal answers (comparing pairs of items) as opposed to when they are asked for numeric evaluations.


\end{itemize}

\subsection{Open problems}
We discuss two sets of open problems, one from the practical perspective and another on the theoretical front.

First, in the paper, we assumed that the number of total questions $\numques$ in a task, the number of gold standard questions $\numgold$, and the threshold $T$ for skipping (or the number and thresholds of the different confidence levels) were provided to the mechanism. While these parameters may be chosen by hand by a system designer based on her own experience, a more principled design of these parameters is an important question. The choices for these parameters may have to be made based on certain tradeoffs. For instance, a higher value of $\numgold$ reduces the variance in the payments but uses more resources in terms of gold standard questions. Or for instance, more number of threshold levels $L$ would increase the amount of information obtained about the workers' beliefs, but also increase the noise in the workers' estimates of her own beliefs. 

A second open problem is the design of inference algorithms that can exploit the specific structure of the skip-based setting. There are several algorithms and theoretical analyses in the literature for aggregating data from multiple workers in the baseline setting~\citepn{raykar2010learning,zhou2012learning, wauthier2011bayesian, chen2013pairwise,kamar2012combining, dawid1979maximum, karger2011iterative,liu2012variational, vempaty2014reliable, zhang2014spectral, ipeirotis2014repeated}. A useful direction of research in the future is to develop algorithms and theoretical guarantees that incorporate information about the workers' confidences. For instance, for the skip-based setting, the missing labels are not missing ``at random'' but are correlated with the difficulty of the task; in the confidence-based setting, we elicit information about the workers' perceived confidence levels. Designing algorithms that can exploit this information judiciously (e.g., via confidence-weighed worker/item constraints in the minimax entropy method of~\citetn{zhou2012learning}) is a useful direction of future research.

\subsection{Conclusions}\label{sec:conclusion}
Despite remarkable progress in machine learning and artificial intelligence, many problems are still not solvable by either humans or machines alone. In recent years, crowdsourcing has emerged as a powerful tool to combine both human and machine intelligence. Crowdsourcing is also a standard means of collecting labeled data for machine learning algorithms. However, crowdsourcing is often plagued with the problem of poor-quality output from workers. 

We designed a reward mechanism for crowdsourcing to ensure collection of high-quality data. Under a very natural ``no-free-lunch'' axiom, we mathematically prove that surprisingly, our mechanism is the only feasible reward mechanism. We further show that among all possible incentive-compatible mechanisms, our ``multiplicative'' mechanism makes the strictly smallest expenditure on spammers. In preliminary experiments, we observe a significant drop in the error rates under this unique mechanism as compared to basic baseline mechanisms, suggesting that our mechanism has the potential to work well in practice. Our mechanisms offer some additional benefits. The pattern of skips or confidence levels of the workers provide a reasonable estimate of the difficulty of each question. In practice, the questions that are estimated to be more difficult may now be delegated to an expert or to more non-expert workers. Secondly, the theoretical guarantees of the mechanism may allow for better post-processing of the data, incorporating the confidence information and improving the overall accuracy. The simplicity of the rules of our mechanisms may facilitate an easier adoption among the workers.

In conclusion, given the uniqueness in theory, simplicity, and good performance observed in practice, we envisage our `multiplicative' mechanisms to be of interest to machine learning researchers and practitioners who use crowdsourcing to collect labeled data.



\section*{Acknowledgements}
The work of Nihar B. Shah was funded in part by a Microsoft Research PhD fellowship. We thank John C. Platt, Christopher J. C. Burges and Christopher Meek for many inspiring discussions. We also thank John C. Platt and Martin J. Wainwright for helping in proof-reading and polishing parts of the manuscript. This work was done when Nihar B. Shah was an intern at Microsoft Research.

\ifjmlr
\else
    \bibliography{crowdsourcing}
\fi

\appendix

\section{Proofs}\label{sec:proofs}
In this section, we prove the claimed theoretical results whose proofs are not included in the main text of the paper. 

The property of incentive-compatibility does not change upon any shift of the mechanism by a constant value or any scaling by a positive constant value. As a result, for the purposes of these proofs, we can assume without loss of generality that $\minpay=0$.

\subsection{Proof of Lemma~\ref{lem:nec_skip}: The Workhorse Lemma}
\label{app:proof_lem_nec_skip}
First we consider the case of ${\numgold}={\numques}$. In the set $\{y_1,\ldots,y_{i-1},y_{i+1},\ldots,y_{\numgold}\}$, for some $(\eta,\gamma) \in \{0,\ldots,{\numgold}-1\}^2$, suppose there are $\eta$ elements with a value $1$, $\gamma$ elements with a value $-1$, and $({\numgold}-1-\eta-\gamma)$ elements with a value $0$. Let us assume for now that $i=\eta+\gamma+1$, $y_1=1,\ldots,y_{\eta}=1,y_{\eta+1}=-1,\ldots,y_{\eta+\gamma}=-1,y_{\eta+\gamma+2}=0,\ldots,y_{\numgold}=0$.

Suppose the worker has confidences $(p_1,\ldots,p_{\eta+\gamma}) \in (T,1]^{\eta+\gamma}$ for the first $(\eta+\gamma)$ questions, a confidence of $q \in (0,1]$ for the next question, and confidences smaller than $T$ for the remaining $({\numgold}-\eta-\gamma-1)$ questions. The mechanism must incentivize the worker to answer the first $(\eta+\gamma)$ questions and skip the last $({\numgold}-\eta-\gamma-1)$ questions; for question $(\eta+\gamma+1)$, it must incentivize the worker to answer if $q>T$ and skip if $q<T$. Supposing the worker indeed attempts the first $(\eta+\gamma)$ questions and skips the last $({\numgold}-\eta-\gamma-1)$ questions, let $\ansbf=\{\ans{1},\ldots,\ans{\eta+\gamma}\} \in \{-1,1\}^{\eta+\gamma}$ denote the the evaluation of the worker's answers to the first $(\eta+\gamma)$ questions. Define quantities $\{r_j\}_{j\in[\eta+\gamma]}$ as $r_j = 1-p_j$ for $j\in \{1,\ldots,\eta\}$, and $r_j=p_j$ for $j \in \{\eta+1,\eta+\gamma\}$. The requirement of incentive compatibility necessitates
\begin{align*}
&q \sum_{\ansbf \in \{-1,1\}^{\eta+\gamma}} \left(f(\ans{1},\ldots,\ans{\eta},-\ans{\eta+1},\ldots,-\ans{\eta+\gamma}, 1,0,\ldots,0) \prod_{j \in [\eta+\gamma]} r_j^{\frac{1-\ans{j}}{2}} (1-r_j)^{\frac{1+\ans{j}}{2}}\right)\nonumber\\
&\qquad+
(1-q) \sum_{\ansbf \in \{-1,1\}^{\eta+\gamma}} \left(f(\ans{1},\ldots,\ans{\eta},-\ans{\eta+1},\ldots,-\ans{\eta+\gamma}, -1,0,\ldots,0) \prod_{j \in [\eta+\gamma]} r_j^{\frac{1-\ans{j}}{2}} (1-r_j)^{\frac{1+\ans{j}}{2}}\right)\nonumber\\
&\overset{q<T}{\underset{q>T}{\lessgtr}}\qquad\quad
\sum_{\ansbf \in \{-1,1\}^{\eta+\gamma}} \left(f(\ans{1},\ldots,\ans{\eta},-\ans{\eta+1},\ldots,-\ans{\eta+\gamma}, 0,0,\ldots,0) \prod_{j \in [\eta+\gamma]} r_j^{\frac{1-\ans{j}}{2}} (1-r_j)^{\frac{1+\ans{j}}{2}}\right)~.
\end{align*}
The left hand side of this expression is the expected payment if the worker chooses to answer question $(\eta+\gamma+1)$, while the right hand side is the expected payment if she chooses to skip it. For any real-valued variable $q$, and for any real-valued constants $a$, $b$ and $c$, \[a q ~\overset{q<c}{\underset{q>c}{\lessgtr}}~b \quad \Rightarrow \quad \ ac=b~.\] As a result,
\begin{align}
&T \sum_{\ansbf \in \{-1,1\}^{\eta+\gamma}} \left(f(\ans{1},\ldots,\ans{\eta},-\ans{\eta+1},\ldots,-\ans{\eta+\gamma}, 1,0,\ldots,0) \prod_{j \in [\eta+\gamma]} r_j^{\frac{1-\ans{j}}{2}} (1-r_j)^{\frac{1+\ans{j}}{2}}\right)\nonumber\\
&\quad+
(1-T) \sum_{\ansbf \in \{-1,1\}^{\eta+\gamma}} \!\! \left(f(\ans{1},\ldots,\ans{\eta},-\ans{\eta+1},\ldots,-\ans{\eta+\gamma}, -1,0,\ldots,0) \prod_{j \in [\eta+\gamma]} r_j^{\frac{1-\ans{j}}{2}} (1-r_j)^{\frac{1+\ans{j}}{2}}\right)\nonumber\\
&\quad-
\sum_{\ansbf \in \{-1,1\}^{\eta+\gamma}} \!\! \left(f(\ans{1},\ldots,\ans{\eta},-\ans{\eta+1},\ldots,-\ans{\eta+\gamma-1}, 0,0,\ldots,0) \prod_{j \in [\eta+\gamma]} r_j^{\frac{1-\ans{j}}{2}} (1-r_j)^{\frac{1+\ans{j}}{2}}\right)=0.\label{eq:lineareqn_skip_proof1}
\end{align}
The left hand side of~\eqref{eq:lineareqn_skip_proof1} represents a polynomial in $(\eta+\gamma)$ variables $\{r_j\}_{j=1}^{\eta+\gamma}$ which evaluates to zero for all values of the variables within a $(\eta+\gamma)$-dimensional solid Euclidean ball. Thus, the coefficients of the monomials in this polynomial must be zero. In particular, the constant term must be zero. The constant term appears when $\ans{j}=1~\forall~j$ in the summations in~\eqref{eq:lineareqn_skip_proof1}. Setting the constant term to zero gives
\begin{align*}
&T f(\ans{1}=1,\ldots,\ans{\eta}=1,-\ans{\eta+1}=-1,\ldots,-\ans{\eta+\gamma}=-1, 1,0,\ldots,0)\nonumber\\
&\qquad+(1-T) f(\ans{1}=1,\ldots,\ans{\eta}=1,-\ans{\eta+1}=-1,\ldots,-\ans{\eta+\gamma}=-1, -1,0,\ldots,0) \nonumber\\
&\qquad\qquad - f(\ans{1}=1,\ldots,\ans{\eta}=1,-\ans{\eta+1}=-1,\ldots,-\ans{\eta+\gamma}=-1, 0,0,\ldots,0) = 0~
\end{align*}
as desired.
Since the arguments above hold for any permutation of the ${\numgold}$ questions, this completes the proof for the case of ${\numgold}={\numques}$.

Now consider the case ${\numgold}<{\numques}$. Let $g:\{-1,0,1\}^{\numques} \rightarrow \mathbb{R}_+$ represent the expected payment given an evaluation of all the ${\numques}$ answers, when the identities of the gold standard questions are unknown. Here, the expectation is with respect to the (uniformly random) choice of the ${\numgold}$ gold standard questions. If $(\ans{1},\ldots,\ans{\numques})\in\{-1,0,1\}^{\numques}$ are the evaluations of the worker's answers to the ${\numques}$ questions then the expected payment is
\beq 
g(\ans{1},\ldots,\ans{\numques}) = \frac{1}{{{\numques} \choose {\numgold}}} \sum_{(i_1,\ldots,i_{\numgold})\subseteq\{1,\ldots,{\numques}\}}  f(\ans{i_1},\ldots,\ans{i_{\numgold}})~.\label{eq:skip_notation_subsample}
\eeq
Notice that when ${\numgold}={\numques}$, the functions $f$ and $g$ are identical.

In the set $\{y_1,\ldots,y_{i-1},y_{i+1},\ldots,y_{\numgold}\}$, for some $(\eta,\gamma) \in \{0,\ldots,{\numgold}-1\}^2$ with $\eta+\gamma < \numgold$, suppose there are $\eta$ elements with a value $1$, $\gamma$ elements with a value $-1$, and $({\numgold}-1-\eta-\gamma)$ elements with a value $0$. Let us assume for now that $i=\eta+\gamma+1$, $y_1=1,\ldots,y_{\eta}=1,y_{\eta+1}=-1,\ldots,y_{\eta+\gamma}=-1,y_{\eta+\gamma+2}=0,\ldots,y_{\numgold}=0$. 

Suppose the worker has confidences $\{p_1,\ldots,p_{\eta+\gamma}\} \in (T,1]^{\eta+\gamma}$ for the first $(\eta+\gamma)$ of the $\numques$ questions, a confidence of $q \in (0,1]$ for the next question, and confidences smaller than $T$ for the remaining $({\numques}-\eta-\gamma-1)$ questions. The mechanism must incentivize the worker to answer the first $(\eta+\gamma)$ questions and skip the last $({\numques}-\eta-\gamma-1)$ questions; for the $(\eta+\gamma+1)^{\rm th}$ question, the mechanism must incentivize the worker to answer if $q>T$ and skip if $q<T$. Supposing the worker indeed attempts the first $(\eta+\gamma)$ questions and skips the last $({\numques}-\eta-\gamma-1)$ questions, let $\ansbf=\{\ans{1},\ldots,\ans{\eta+\gamma}\} \in \{-1,1\}^{\eta+\gamma}$ denote the the evaluation of the worker's answers to the first $(\eta+\gamma)$ questions. Define quantities $\{r_j\}_{j \in [\eta+\gamma]}$ as $r_j = 1-p_j$ for $j\in \{1,\ldots,\eta\}$, and $r_j=p_j$ for $j \in \{\eta+1,\eta+\gamma\}$. The requirement of incentive compatibility necessitates
\begin{align}
&q \sum_{\ansbf \in \{-1,1\}^{\eta+\gamma}} \left(g(\ans{1},\ldots,\ans{\eta},-\ans{\eta+1},\ldots,-\ans{\eta+\gamma}, 1,0,\ldots,0) \prod_{j \in [\eta+\gamma]} r_j^{\frac{1-\ans{j}}{2}} (1-r_j)^{\frac{1+\ans{j}}{2}}\right)\nonumber\\
&\quad+
(1-q) \sum_{\ansbf \in \{-1,1\}^{\eta+\gamma}} \!\! \left(g(\ans{1},\ldots,\ans{\eta},-\ans{\eta+1},\ldots,-\ans{\eta+\gamma}, -1,0,\ldots,0) \prod_{j \in [\eta+\gamma]} r_j^{\frac{1-\ans{j}}{2}} (1-r_j)^{\frac{1+\ans{j}}{2}}\right)\nonumber\\
&\overset{q<T}{\underset{q>T}{\lessgtr}}\qquad
\sum_{\ansbf \in \{-1,1\}^{\eta+\gamma}} \!\! \left(g(\ans{1},\ldots,\ans{\eta},-\ans{\eta+1},\ldots,-\ans{\eta+\gamma}, 0,0,\ldots,0) \prod_{j \in [\eta+\gamma]} r_j^{\frac{1-\ans{j}}{2}} (1-r_j)^{\frac{1+\ans{j}}{2}}\right)~.\label{eq:skip_necsuf_1}
\end{align}
Again, applying the fact that for any real-valued variable $q$ and for any real-valued constants $a$, $b$ and $c$, $a q ~\overset{q<c}{\underset{q>c}{\lessgtr}}~b \quad \Rightarrow  \ ac=b$, we get that
\begin{align}
&T g(\ans{1}=1,\ldots,\ans{\eta}=1,-\ans{\eta+1}=-1,\ldots,-\ans{\eta+\gamma}=-1, 1,0,\ldots,0)\nonumber\\
&\qquad+(1-T) g(\ans{1}=1,\ldots,\ans{\eta}=1,-\ans{\eta+1}=-1,\ldots,-\ans{\eta+\gamma}=-1, -1,0,\ldots,0) \nonumber\\
&\qquad\qquad - g(\ans{1}=1,\ldots,\ans{\eta}=1,-\ans{\eta+1}=-1,\ldots,-\ans{\eta+\gamma}=-1, 0,0,\ldots,0) = 0~\label{eq:skip_necsuf_2}.
\end{align}

The proof now proceeds via induction on the quantity $({\numgold}-\eta-\gamma-1)$, i.e., on the number of skipped questions in $\{y_1,\ldots,y_{i-1},y_{i+1},\ldots,y_{\numgold}\}$. We begin with the case of $({\numgold}-\eta-\gamma-1)={\numgold}-1$ which implies $\eta=\gamma=0$. In this case~\eqref{eq:skip_necsuf_2} simplifies to
\begin{align*}
&T g(1,0,\ldots,0) + (1-T) g(-1,0,\ldots,0) = g(0,0,\ldots,0)~.
\end{align*}
Applying the expansion of function $g$ in terms of function $\payfn$ from~\eqref{eq:skip_notation_subsample} gives
\begin{align*}
&T \left(c_1 \payfn(1,0,\ldots,0)+c_2f(0,0,\ldots,0)\right) + (1-T) \left(c_1 \payfn(-1,0,\ldots,0)+c_2 \payfn(0,0,\ldots,0)\right) \nonumber\\
&\qquad\qquad\qquad\qquad\qquad\qquad\qquad\qquad\qquad\qquad\qquad= \left(c_1 f(0,0,\ldots,0)+c_2 \payfn(0,0,\ldots,0)\right)
\end{align*}
for constants $c_1>0$ and $c_2>0$ that respectively denote the probabilities that the first question is picked and not picked in the set of ${\numgold}$ gold standard questions. Cancelling out the common terms on both sides of the equation, we get the desired result
\begin{align*}
&T \payfn(1,0,\ldots,0) + (1-T) \payfn(-1,0,\ldots,0) = \payfn(0,0,\ldots,0)~.
\end{align*}
Next, we consider the case when $({\numgold}-\eta-\gamma-1)$ questions are skipped in the gold standard, and assume that the result is true when more than $({\numgold}-\eta-\gamma-1)$ questions are skipped in the gold standard. In~\eqref{eq:skip_necsuf_2}, the functions $g$ decompose into a sum of the constituent $f$ functions. These constituent functions $f$ are of two types: the first where all of the first $(\eta+\gamma+1)$ questions are included in the gold standard, and the second where one or more of the first $(\eta+\gamma+1)$ questions are not included in the gold standard. The second case corresponds to situations where there are more than $({\numgold}-\eta-\gamma-1)$ questions skipped in the gold standard and hence satisfies our induction hypothesis. The terms corresponding to these functions thus cancel out in the expansion of~\eqref{eq:skip_necsuf_2}. The remainder comprises only evaluations of function $f$ for arguments in which the first $(\eta+\gamma+1)$ questions are included in the gold standard: since the last $(N-\eta-\gamma-1)$ questions are skipped by the worker, the remainder evaluates to
\begin{align*}
&T c_3 \payfn(y_1,\ldots,y_{\eta+\gamma},1,0,\ldots,0) + (1-T) c_3 \payfn(y_1,\ldots,y_{\eta+\gamma},-1,0,\ldots,0) \nonumber\\
&\qquad\qquad\qquad\qquad\qquad\qquad\qquad\qquad\qquad\qquad\qquad= c_3 f(y_1,\ldots,y_{\eta+\gamma},0,0,\ldots,0)
\end{align*}
for some constant $c_3>0$. Dividing throughout by $c_3$ gives the desired result.

Finally, the arguments above hold for any permutation of the first ${\numgold}$ questions, thus completing the proof.

\subsection{Proof of Theorem~\ref{thm:mechanism_working_confidence}: Working of Algorithm~\ref{algo:incentive_confidence}}
\label{app:proof_incentive_confidence}
We first state three properties that the constants $\{\alpha_\elll\}_{\elll = -L}^{L}$ defined in  Algorithm~\ref{algo:incentive_confidence} must satisfy. We will use these properties subsequently in the proof of Theorem~\ref{thm:mechanism_working_confidence}.

\begin{lemma}\label{lem:working_confidence_0}
For every $\elll \in \{0,\ldots,L-1\}$
\beq
T_{\elll+1} \alpha_{\elll+1} + (1-T_{\elll+1})\alpha_{-({\elll+1})} = T_{\elll+1}\alpha_{\elll} + (1-T_{\elll+1})\alpha_{-\elll}~,\label{eq:lem_working_confidence_0_statement1}
\eeq
and
\beq
S_{\elll+1} \alpha_{\elll+1} + (1-S_{\elll+1})\alpha_{-({\elll+1})} = \alpha_0=1~.\label{eq:lem_working_confidence_0_statement2}
\eeq
\end{lemma}

\begin{lemma}\label{lem:working_confidence_1}
$\alpha_L>\alpha_{L-1}>\cdots>\alpha_{-L}=0$.
\end{lemma}

\begin{lemma}\label{lem:working_confidence_2}
For any $m \in \{1,\ldots,L\}$, any $p>T_m$ and any $z<m$,
\beq
p\alpha_m + (1-p)\alpha_{-m} > p\alpha_z + (1-p)\alpha_{-z}~,\label{eq:lem_working_confidence_2_statement1}
\eeq
and for any $m \in \{0,\ldots,L-1\}$, any $p<T_{m+1}$ and any $z>m$,
\beq
p\alpha_m + (1-p)\alpha_{-m} > p\alpha_z + (1-p)\alpha_{-z}~. \label{eq:lem_working_confidence_2_statement2}
\eeq
\end{lemma}

The proof of these results are available at the end of this subsection. Assuming these lemmas hold, we will now complete the proof of Theorem~\ref{thm:mechanism_working_confidence}.

The choice of $\alpha_{-L}=0$ made in Algorithm~\ref{algo:incentive_confidence} ensures that the payment is zero whenever any answer in the gold standard evaluates to $-L$. This choice ensures that the no-free-lunch condition is satisfied. One can easily verify that the payment lies in the interval $[0,\maxpay]$. It remains to prove that the proposed mechanism is incentive-compatible.


Define $E = (\epsilon_1,\ldots,\epsilon_{\numgold}) \in \{-1,1\}^{\numgold}$ and $E_{\backslash 1}=(\epsilon_2,\ldots,\epsilon_{\numgold})$.
Suppose the worker has confidences $p_1,\ldots,p_{\numques}$ for her ${\numques}$ answers. For some $(s(1),\ldots,s({\numques})) \in \{0,\ldots,L\}^{\numques}$ suppose $p_i \in (T_{s(i)},\ T_{s(i)+1})~\forall\ i\in\{1,\ldots,{\numques}\}$, i.e., $s(1),\ldots,s({\numques})$ are the correct confidence-levels for her answers. Consider any other set of confidence-levels $s'(1),\ldots,s'({\numques})$. When the mechanism of Algorithm~\ref{algo:incentive_confidence} is employed, the expected payment (from the point of view of the worker) on selecting confidence-levels $s(1),\ldots,s({\numques})$ is
\begin{align} 
\mathbb{E}\textrm{[Pay]} =& \frac{1}{{{\numques} \choose {\numgold}}} \sum_{\substack{(j_1,...,j_{\numgold})\\\subseteq\{1,\ldots,{\numques}\}}} \sum_{E\in\{-1,1\}^{\numgold}} \prod_{i=1}^{{\numgold}} \alpha_{\epsilon_i s(j_i)} (p_{j_i})^{\frac{1+\epsilon_i}{2}} (1-p_{j_i})^{\frac{1-\epsilon_i}{2}} \label{eq:mechanism_working_confidence_1}\\
=& \frac{1}{{{\numques} \choose {\numgold}}} \sum_{\substack{(j_1,...,j_{\numgold})\\\subseteq\{1,\ldots,{\numques}\}}} \sum_{E_{\backslash 1}\in\{-1,1\}^{{\numgold}-1}}\!\! \left(p_{j_1}\alpha_{s({j_1})} + (1-p_{j_1}) \alpha_{-s({j_1})}\right) \prod_{i=2}^{{\numgold}} \alpha_{\epsilon_i s(j_i)} (p_{j_i})^{\frac{1+\epsilon_i}{2}} (1-p_{j_i})^{\frac{1-\epsilon_i}{2}}\\
&\vdots\nonumber\\
=& \frac{1}{{{\numques} \choose {\numgold}}} \sum_{\substack{(j_1,...,j_{\numgold})\\\subseteq\{1,\ldots,{\numques}\}}} \prod_{i=1}^{{\numgold}} \left(p_{j_i}\alpha_{s({j_i})}+ (1-p_{j_i}) \alpha_{-s({j_i})}\right)\label{eq:mechanism_working_confidence_2}\\
>&\frac{1}{{{\numques} \choose {\numgold}}} \sum_{\substack{(j_1,...,j_{\numgold})\\\subseteq\{1,\ldots,{\numques}\}}} \prod_{i=1}^{{\numgold}} \left(p_{j_i}\alpha_{s'({j_i})} + (1-p_{j_i}) \alpha_{-s'({j_i})}\right)
\end{align}
which is the expected payment under any other set of confidence-levels $s'(1),\ldots,s'({\numques})$. The last inequality is a consequence of Lemma~\ref{lem:working_confidence_2}.

An argument similar to the above also proves that for any $m\in\{1,\ldots,L\}$, if allowed to choose between only skipping and confidence-level $m$, the worker is incentivized to choose confidence-level $m$ over skip if her confidence is greater $S_m$, and choose skip over level $m$ if if her confidence is smaller than $S_m$. Finally, from Lemma~\ref{lem:working_confidence_1} we have $\alpha_L>\cdots>\alpha_{-L}=0$. It follows that the expected payment~\eqref{eq:mechanism_working_confidence_2} is strictly increasing in each of the values $p_1,\ldots,p_{\numques}$. Thus the worker is incentivized to report the answer that she thinks is most likely to be correct. 

\subsubsection{Proof of Lemma~\ref{lem:working_confidence_0}}
Algorithm~\ref{algo:incentive_confidence} states that $\alpha_{-\elll} = \frac{1 - \alpha_{\elll} S_\elll}{1-S_\elll}$ for all $\elll \in [L]$. A simple rearrangement of the terms in this expression gives~\eqref{eq:lem_working_confidence_0_statement2}.

Towards the goal of proving~\eqref{eq:lem_working_confidence_0_statement1}, we will first prove an intermediate result: 
\begin{equation}
\alpha_\elll>1>\alpha_{-\elll}~\forall~\elll \in \{L,\ldots,1\}~.
\label{eq:alphaelll_grtr_alphaminuselll}
\end{equation}
The proof proceeds via an induction on $\elll \in \{L,\ldots,2\}$. The case of $\elll=1$ will be proved separately. The induction hypothesis involves two claims: $\alpha_{\elll} > 1 > \alpha_{-\elll}$ and $T_{\elll} \alpha_{\elll} + (1-T_{\elll}) \alpha_{-\elll} > 1$. The base case is $\elll=L$ for which we know that $\alpha_L = \frac{1}{S_L} >1 > 0 = \alpha_{-L}$ and $T_{\elll} \alpha_{\elll} + (1-T_{\elll}) \alpha_{-\elll} = \frac{T_\elll}{S_\elll}>1$. Now suppose that the induction hypothesis is true for $(\elll+1)$. Rearranging the terms in the expression defining $\alpha_\elll$ in Algorithm~\ref{algo:incentive_confidence} and noting that $1>T_{\elll+1}>S_\elll$, we get
\begin{align}
\alpha_{\elll} &= \frac{(1-S_\elll) \left( T_{\elll+1}\alpha_{\elll+1}+(1-T_{\elll+1})\alpha_{-(\elll+1)}\right)-(1-T_{\elll+1})}{(1-S_\elll)-(1-T_{\elll+1})} \label{eq:lem_working_confidence_proof1}\\
&> \frac{(1-S_\elll)-(1-T_{\elll+1})}{(1-S_\elll)-(1-T_{\elll+1})}
\\
&=1~.
\end{align}
From~\eqref{eq:lem_working_confidence_0_statement2} we see that the value $1$ is a convex combination of $\alpha_\elll$ and $\alpha_{-\elll}$. Since $\alpha_\elll>1$ and $S_{\elll} \in (0,1)$, it must be that $\alpha_{-\elll}<1$. Furthermore, since $T_{\elll}>S_\elll$ we get
\begin{align}
T_\elll \alpha_\elll + (1-T_\elll) \alpha_{-\elll} &> S_\elll \alpha_\elll + (1-S_\elll) \alpha_{-\elll} \\
&= 1~.
\end{align}
This proves the induction hypothesis. Let us now consider $\elll = 1$. If $L=1$ then we have $\alpha_L = \frac{1}{S_L} >1 > 0 = \alpha_{-L}$ and we are done. If $L>1$ then we have already proved that $\alpha_{2} > 1 > \alpha_{-2}$ and $T_{2} \alpha_{2} + (1-T_{2}) \alpha_{-2} > 1$. An argument identical to~\eqref{eq:lem_working_confidence_proof1} onwards proves that $\alpha_1 > 1 > \alpha_{-1}$.

Now that we have proved $\alpha_\elll > \alpha_{-\elll} \forall ~\elll \in [L]$, we can rewrite the expression defining $\alpha_{-\elll}$ in Algorithm~\ref{algo:incentive_confidence} as
\begin{align}
S_\elll = \frac{1 - \alpha_{-\elll}}{\alpha_\elll - \alpha_{-\elll}}~.
\end{align}
Substituting this expression for $S_\elll$ in the definition of $\alpha_\elll$ in Algorithm~\ref{algo:incentive_confidence} and making some simple rearrangements gives the desired result~\eqref{eq:lem_working_confidence_0_statement1}.

\subsubsection{Proof of Lemma~\ref{lem:working_confidence_1}}
We have already shown~\eqref{eq:alphaelll_grtr_alphaminuselll} in the proof of Lemma~\ref{lem:working_confidence_0} above that $\alpha_\elll > 1 > \alpha_{-\elll}~\forall\ \elll \in [L]$.

Next we will show that $\alpha_{\elll+1} > \alpha_{\elll}$ and $\alpha_{-(\elll+1)} < \alpha_{-\elll}$~$\forall~\elll \geq 0$. First consider $\elll=0$, for which Algorithm~\ref{algo:incentive_confidence} sets $\alpha_0=1$, and we have already proved that $\alpha_1 > 1 > \alpha_{-1}$.

Now consider some $\elll>0$. Observe that since $S_\elll \alpha_\elll +(1-S_\elll) \alpha_{-\elll}=1$ (Lemma~\ref{lem:working_confidence_0}), $S_{\elll+1}>S_\elll$ and $\alpha_\elll>\alpha_{-\elll}$, it must be that \beq S_{\elll+1} \alpha_\elll +(1-S_{\elll+1}) \alpha_{-\elll} >1~.\label{eq:mechanism_working_confidence_posgtneg6}\eeq
From Lemma~\ref{lem:working_confidence_0}, we also have 
 \beq S_{\elll+1} \alpha_{\elll+1} +(1-S_{\elll+1}) \alpha_{-(\elll+1)} =1~.\label{eq:mechanism_working_confidence_posgtneg7}\eeq
Subtracting~\eqref{eq:mechanism_working_confidence_posgtneg6} from~\eqref{eq:mechanism_working_confidence_posgtneg7} we get
  \beq S_{\elll+1} (\alpha_{\elll+1}-\alpha_\elll) +(1-S_{\elll+1}) (\alpha_{-(\elll+1)}-\alpha_{-\elll})<0~.\label{eq:mechanism_working_confidence_posgtneg8}\eeq
From Lemma~\ref{lem:working_confidence_0}  we also have 
 \beq T_{\elll+1} \alpha_{\elll+1} +(1-T_{\elll+1}) \alpha_{-(\elll+1)} =T_{\elll+1} \alpha_{\elll} +(1-T_{\elll+1}) \alpha_{-\elll}\label{eq:mechanism_working_confidence_posgtneg9}\eeq
  \beq \Rightarrow\quad T_{\elll+1} (\alpha_{\elll+1}-\alpha_\elll) +(1-T_{\elll+1}) (\alpha_{-(\elll+1)}-\alpha_{-\elll})=0~.\label{eq:mechanism_working_confidence_posgtneg10}\eeq
Subtracting~\eqref{eq:mechanism_working_confidence_posgtneg8} from~\eqref{eq:mechanism_working_confidence_posgtneg10} gives
  \beq (T_{\elll+1} -S_{\elll+1}  ) [( \alpha_{\elll+1}-\alpha_\elll) + (\alpha_{-\elll}-\alpha_{-(\elll+1)})]>0~.\label{eq:mechanism_working_confidence_posgtneg11}\eeq
Since $T_{\elll+1}>S_{\elll+1}$ by definition, it must be that
 \beq
  \alpha_{\elll+1}-\alpha_\elll > \alpha_{-(\elll+1)}-\alpha_{-\elll}~.\label{eq:mechanism_working_confidence_posgtneg12}
 \eeq
Now, rearranging the terms in~\eqref{eq:mechanism_working_confidence_posgtneg9} gives
 \beq
  (\alpha_{\elll+1}-\alpha_\elll)T_{\elll+1} =- (\alpha_{-(\elll+1)}-\alpha_{-\elll})(1-T_{\elll+1})~.\label{eq:mechanism_working_confidence_posgtneg13}
\eeq
Since $T_{\elll+1}\in(0,1)$, it follows that  the terms $(\alpha_{\elll+1}-\alpha_\elll)$ and $(\alpha_{-(\elll+1)}-\alpha_{-\elll})$ have opposite signs. Using~\eqref{eq:mechanism_working_confidence_posgtneg12} we conclude that
$\alpha_{\elll+1}-\alpha_\elll>0$ and $\alpha_{-(\elll+1)}-\alpha_{-\elll}<0$.

\subsubsection{Proof of Lemma~\ref{lem:working_confidence_2}}
Let us first prove~\eqref{eq:lem_working_confidence_2_statement1}. First consider the case $z=m-1$. From Lemma~\ref{lem:working_confidence_0} we know that
\bea
T_m\alpha_{m-1} + (1-T_m)\alpha_{-(m-1)}&=&T_m\alpha_m + (1-T_m)\alpha_{-m}\nonumber\\
\Rightarrow\qquad
0&=&T_m(\alpha_m-\alpha_{m-1}) + T_m(\alpha_{-(m-1)}-\alpha_{-m}) -(\alpha_{-(m-1)}-\alpha_{-m})\nonumber\\
&< &p(\alpha_m-\alpha_{m-1}) + p(\alpha_{-(m-1)}-\alpha_{-m}) -(\alpha_{-(m-1)}-\alpha_{-m})~,\label{eq:lem_working_confidence_2_1}
\eea
where~\eqref{eq:lem_working_confidence_2_1} is a consequence of $p>T_m$ and Lemma~\ref{lem:working_confidence_1}. A simple rearrangement of the terms in~\eqref{eq:lem_working_confidence_2_1} gives~\eqref{eq:lem_working_confidence_2_statement1}. Now, for any $z<m$, recursively apply this result to get
\bea
p\alpha_m + (1-p)\alpha_{-m} &>& p\alpha_{m-1} + (1-p)\alpha_{-(m-1)}\nonumber\\
&>& p\alpha_{m-2} + (1-p)\alpha_{-(m-2)}\nonumber\\
&\vdots&\nonumber\\
&>& p\alpha_z + (1-p)\alpha_{-z}~.\nonumber
\eea

Let us now prove~\eqref{eq:lem_working_confidence_2_statement2}. We first consider the case $z=m+1$. From Lemma~\ref{lem:working_confidence_0} we know that
\bea
T_{m+1}\alpha_m + (1-T_{m+1})\alpha_{-m} &=& T_{m+1}\alpha_{m+1} + (1-T_{m+1})\alpha_{-(m+1)}~\nonumber\\
\Rightarrow\qquad
0&=&T_{m+1}(\alpha_{m+1}-\alpha_m) + T_{m+1}(\alpha_{-m}-\alpha_{-(m+1)}) -(\alpha_{-m}-\alpha_{-(m+1)})\nonumber\\
&> &p(\alpha_{m+1}-\alpha_m) + p(\alpha_{-m}-\alpha_{-(m+1)}) -(\alpha_{-m}-\alpha_{-(m+1)})~,\label{eq:lem_working_confidence_2_2}
\eea
where~\eqref{eq:lem_working_confidence_2_2} is a consequence of $p<T_{m+1}$ and Lemma~\ref{lem:working_confidence_1}. A simple rearrangement of the terms in~\eqref{eq:lem_working_confidence_2_2} gives~\eqref{eq:lem_working_confidence_2_statement2}. For any $z>m$, applying this result recursively gives
\bea
p\alpha_m + (1-p)\alpha_{-m} &>& p\alpha_{m+1} + (1-p)\alpha_{-(m+1)}\nonumber\\
&>& p\alpha_{m+2} + (1-p)\alpha_{-(m+2)}\nonumber\\
&\vdots&\nonumber\\
&>& p\alpha_z + (1-p)\alpha_{-z}~.\nonumber
\eea

\subsection{Proof of Theorem~\ref{thm:unique_confidence}: Uniqueness of Algorithm~\ref{algo:incentive_confidence}}
\label{app:proof_unique_confidence}
We will first define one additional piece of notation. Let $g:\{-L,\ldots,L\}^{\numques} \rightarrow \mathbb{R}_+$ denote the expected payment given an evaluation of the ${\numques}$ answers, where the expectation is with respect to the (uniformly random) choice of the ${\numgold}$ gold standard questions. If $(\ans{1},\ldots,\ans{\numques})\in\{-L,\ldots,L\}^{\numques}$ are the evaluations of the worker's answers to the ${\numques}$ questions then the expected payment is
\beq 
g(\ans{1},\ldots,\ans{\numques}) = \frac{1}{{{\numques} \choose {\numgold}}} \sum_{(i_1,\ldots,i_{\numgold})\subseteq\{1,\ldots,{\numques}\}}  f(\ans{i_1},\ldots,\ans{i_{\numgold}})~.
\label{eq:confidence_notation_subsample}
\eeq
Notice that when ${\numgold}={\numques}$, the functions $f$ and $g$ are identical.

The proof of uniqueness is based on a certain condition necessitated by incentive-compatibility stated in the form of Lemma~\ref{lem:nec_confidence} below. Note that this lemma does \textit{not} require the generalized-no-free-lunch condition, and may be of independent interest.

\begin{lemma}
Any incentive-compatible mechanism must satisfy, for every question $i \in \{1,\ldots,{\numgold}\}$, every\\ $(y_1,\ldots,y_{i-1},y_{i+1},\ldots,y_{\numgold} ) \in \{-L,\ldots,L\}^{{\numgold}-1}$, and every $m \in \{1,\ldots,L\}$,
\begin{align}
&T_m f(y_1,\ldots,y_{i-1},m,y_{i+1},\ldots,y_{\numgold}) + (1-T_m)f(y_1,\ldots,y_{i-1},-m,y_{i+1},\ldots,y_{\numgold})\nonumber\\
&\quad = T_m f(y_1,\ldots,y_{i-1},m-1,y_{i+1},\ldots,y_{\numgold}) + (1-T_m) f(y_1,\ldots,y_{i-1},-(m-1),y_{i+1},\ldots,y_{\numgold})\label{eq:nec_confidence_T}
\end{align}
and
\begin{align}
&S_m f(y_1,\ldots,y_{i-1},m,y_{i+1},\ldots,y_{\numgold}) + (1-S_m)f(y_1,\ldots,y_{i-1},-m,y_{i+1},\ldots,y_{\numgold})\nonumber\\
&\qquad = f(y_1,\ldots,y_{i-1},0,y_{i+1},\ldots,y_{\numgold})~. \label{eq:nec_confidence_S}
\end{align}
\label{lem:nec_confidence}
\end{lemma}
Note that~\eqref{eq:nec_confidence_T} and~\eqref{eq:nec_confidence_S} coincide when $m=1$, since $T_1=S_1$ by definition.

We will first prove that any incentive compatible mechanism that satisfies the no-free-lunch condition must give a zero payment when one or more questions are selected with a confidence $L$ and turn out to be incorrect. Let us assume for now that in the ${\numgold}$ questions in the gold standard, the first question is answered incorrectly with a confidence of $L$, the next $({\numgold}-1-S)$ questions are answered by the worker and have arbitrary evaluations, and the remaining $S$ questions are skipped. The proof proceeds by an induction on $S$. If $S={\numgold}-1$, the only attempted question is the first question and this is incorrect with confidence $L$. The no-free-lunch condition necessitates a zero payment in this case, thus satisfying the base case of our induction hypothesis. Now we prove the hypothesis for some $S$ under the assumption that the hypothesis is true for every $S'>S$. From Lemma~\ref{lem:nec_skip} with $m=1$, we have
\begin{align}
&T_1 f(-L,y_2,\ldots,y_{{\numgold}-S-1},1,0,\ldots,0) + (1-T_1) f(-L,y_2,\ldots,y_{{\numgold}-S-1},-1,0,\ldots,0)\nonumber\\
&\qquad\qquad = T_1 f(-L,y_2,\ldots,y_{{\numgold}-S-1},0,0,\ldots,0)~+(1-T_1)f(-L,y_2,\ldots,y_{{\numgold}-S-1},0,0,\ldots,0)\nonumber\\
&\qquad\qquad = f(-L,y_2,\ldots,y_{{\numgold}-S-1},0,0,\ldots,0)\nonumber\\
&\qquad\qquad =0~,\label{eq:unique_confidence_thm_eq1}
\end{align}
where the final equation~\eqref{eq:unique_confidence_thm_eq1} is a consequence of our induction hypothesis given the fact that $f(-L,y_2,\ldots,y_{{\numgold}-S-1},0,0,\ldots,0)$ corresponds to the case when the last $(S+1)$ questions are skipped and the first question is answered incorrectly with confidence $L$. Now, since the payment $f$ must be non-negative and since $T \in (0,1)$, it must be that 
\beq f(-L,y_2,\ldots,y_{{\numgold}-S-1},1,0,\ldots,0)=0 \label{eq:unique_confidence_thm_eq2} \eeq and 
\beq f(-L,y_2,\ldots,y_{{\numgold}-S-1},-1,0,\ldots,0)=0~.\eeq Repeatedly applying the same argument to $m=2,\ldots,L$ gives that for every value of $m$, it must be that $f(-L,y_2,\ldots,y_{{\numgold}-S-1},m,0,\ldots,0)= f(-L,y_2,\ldots,y_{{\numgold}-S-1},-m,0,\ldots,0)=0$. This completes the proof of our induction hypothesis. Observe that each of the aforementioned arguments hold for any permutation of the ${\numgold}$ questions, thus proving the necessity of zero payment when any one or more answers are incorrect. 

We will now prove that when no answers in the gold standard are incorrect with confidence $L$, the payment must be of the form described in Algorithm~\ref{algo:incentive_skip}. Let $\kappa$ denote the payment when all ${\numgold}$ questions in the gold standard are skipped, i.e., \[ \kappa = f(0,\ldots,0)~.\] 
Now consider any $S \in \{0,\ldots,{\numgold}-1\}$ and any $(y_1,\ldots,y_{{\numgold}-S-1},m)\in \{-L,\ldots,L\}^{{\numgold}-S}$. The payments $\{f(y_1,\ldots,y_{{\numgold}-S-1},m,0,\ldots,0)\}_{m =-L}^{L}$ must satisfy the $(2L-1)$ linear constraints arising out of Lemma~\ref{lem:nec_confidence} and must also satisfy $f(y_1,\ldots,y_{{\numgold}-S-1},-L,0,\ldots,0) =0$. This comprises a total of $2L$ linearly independent constraints on the $(2L+1)$ values $\{f(y_1,\ldots,y_{{\numgold}-S-1},m,0,\ldots,0)\}_{m=-L}^{L}$. The only set of solutions that meet these constraints are
\[
f(y_1,\ldots,y_{{\numgold}-S-1},m,0,\ldots,0) = \alpha_m f(y_1,\ldots,y_{{\numgold}-S-1},0,0,\ldots,0)~,
\]
where the constants $\{\alpha_m\}_{m=-L}^{L}$ are as specified in Algorithm~\ref{algo:incentive_confidence}.
Applying this argument ${\numgold}$ times, starting from $S=0$ to $S={\numgold}-1$, gives
\[
f(y_1,\ldots,y_{{\numgold}}) = \kappa  \prod_{j=1}^{{\numgold}} \alpha_{y_j}~.
\]
Finally, the budget requirement necessitates $\maxpay = \kappa \left(\alpha_L\right)^\numgold$, which mandates the value of $\kappa$ to be $\maxpay \left(\frac{1}{\alpha_L}\right)^\numgold$. This is precisely the mechanism described in Algorithm~\ref{algo:incentive_confidence}.

\subsection{Proof of Lemma~\ref{lem:nec_confidence}: Necessary condition for any incentive-compatible mechanism}
\label{app:proof_lemma_necsuf_confidence}
First consider the case of ${\numgold}={\numques}$. For every $j\in \{1,\ldots,i-1,i+1,\ldots,{\numgold}\}$, define \[r_j =\begin{cases} 1-p_j &\mbox{if}\quad y_j\geq 0\\p_j &\mbox{if}\quad y_j<0~.\end{cases}\] 
Define $E_{\backslash i} = \{\epsilon_1,\ldots,\epsilon_{i-1},\epsilon_{i+1},\ldots,\epsilon_{\numgold}\}$. For any $\elll \in \{-L,\ldots,L\}$ let $\Lambda_\elll \in \mathbb{R}_+$ denote the expected payment (from the worker's point of view) when her answer to the $i^\textrm{th}$ question evaluates to $\elll$:
\bea
\Lambda_{\elll} \!\!&=&\!\!\!\!\!\sum_{E_{\backslash i} \in \{-1,1\}^{{\numgold}-1}} \!\! \left(\!\! f(y_1\epsilon_1,\ldots,y_{i-1}\epsilon_{i-1},\elll,y_{i+1}\epsilon_{i+1},\ldots,y_{\numgold}\epsilon_{\numgold}) \!\!\!\! \prod_{j \in [{\numgold}]\backslash\{i\}} r_j^{\frac{1-\epsilon_j}{2}} \!\! (1-r_j)^{\frac{1+\epsilon_j}{2}}\right).
\label{eq:nec_confidence1}
\eea

Consider a worker who has confidences $\{p_1,\ldots,p_{i-1},p_{i+1},\ldots,p_{\numgold}\} \in (0,1)^{{\numgold}-1}$ for questions $\{1,\ldots,i-1,i+1,\ldots,{\numgold}\}$ respectively, and for question $i$ suppose she has a confidence of $q\in(T_{m-1},T_{m+1})$. For question $i$, we must incentivize the worker to select confidence-level $m$ if $q > T_m$, and to select $(m-1)$ if $q<T_m$. This necessitates
\beq
q \Lambda_{m} + (1-q) \Lambda_{-m} \overset{q<T_m}{\underset{q>T_m}{\lessgtr}} q \Lambda_{m-1} + (1-q) \Lambda_{-(m-1)}~.\label{eq:nec_confidence2}
\eeq
Also, for question $i$, the requirement of level $m$ having a higher incentive as compared to skipping when $q>S_m$ and vice versa when $q<S_m$ necessitates
\beq
q \Lambda_{m} + (1-q) \Lambda_{-m} \overset{q<S_m}{\underset{q>S_m}{\lessgtr}} \Lambda_{0}~.\label{eq:nec_confidence2S}
\eeq
Now, note that for any real-valued variable $q$, and for any real-valued constants $a$, $b$ and $c$, \[aq~\overset{q<c}{\underset{q>c}{\lessgtr}}~b \quad \Rightarrow \quad ac=b~.\]
Applying this fact to~\eqref{eq:nec_confidence2} and~\eqref{eq:nec_confidence2S} gives
\beq
(T_m \Lambda_{m} + (1-T_m)\Lambda_{-{m}}) - (T_m \Lambda_{{m-1}} + (1-T_m) \Lambda_{-{(m-1)}})=0~,
\label{eq:nec_confidence4}
\eeq
\beq
(S_m \Lambda_{m} + (1-S_m)\Lambda_{-{m}}) - \Lambda_{{0}}=0~.
\label{eq:nec_confidence4S}
\eeq
From the definition of $\Lambda_\elll$ in~\eqref{eq:nec_confidence1}, we see that the left hand sides of~\eqref{eq:nec_confidence4} and~\eqref{eq:nec_confidence4S} are both polynomials in $(\numgold-1)$ variables $\{r_j\}_{j \in [{\numgold}]\backslash\{i\}}$ and take a value of zero for all values of the variables in a $(\numgold-1)$-dimensionall solid ball. Thus, each of the coefficients (of the monomials) in both polynomials must be zero, and in particular, the constant terms must also be zero. Observe that in both these polynomials, the constant term arises only when $\epsilon_j=1~\forall~j\in [{\numgold}]\backslash\{i\}$ (which makes the exponent of $r_j$ to be $0$ and that of $(1-r_j)$ to be $1$). Thus, setting the constant term to zero in the two polynomials results in
\begin{align}
&T_m f(y_1,\ldots,y_{i-1},m,y_{i+1},\ldots,y_{\numgold}) + (1-T_m)f(y_1,\ldots,y_{i-1},-m,y_{i+1},\ldots,y_{\numgold})\nonumber\\
&\quad = T_m f(y_1,\ldots,y_{i-1},m-1,y_{i+1},\ldots,y_{\numgold}) + (1-T_m) f(y_1,\ldots,y_{i-1},-(m-1),y_{i+1},\ldots,y_{\numgold})\label{eq:nec_confidence_4}
\end{align}
and
\begin{align}
&S_m f(y_1,\ldots,y_{i-1},m,y_{i+1},\ldots,y_{\numgold}) + (1-S_m)f(y_1,\ldots,y_{i-1},-m,y_{i+1},\ldots,y_{\numgold})\nonumber\\
&\qquad = f(y_1,\ldots,y_{i-1},0,y_{i+1},\ldots,y_{\numgold}) \label{eq:nec_confidence_4S}
\end{align}
thus proving the claim for the case of $\numgold = \numques$.

Now consider the case when ${\numgold}<{\numques}$. In order to simplify notation, let us assume $i=1$ without loss of generality (since the arguments presented hold for any permutation of the questions). Suppose a worker's answers to questions $\{2,\ldots,{\numgold}\}$ evaluate to $(y_2,\ldots,y_{\numgold}) \in \{-L,\ldots,L\}^{{\numgold}-1}$, and further suppose that the worker skips the remaining $({\numques}-{\numgold})$ questions. By going through arguments identical to those for ${\numgold}={\numques}$, but with $f$ replaced by $g$, we get the necessity of
\begin{align}
&T_m g(m,y_2,\ldots,y_{\numgold},0,\ldots,0) + (1-T_m)g(-m,y_2,\ldots,y_{\numgold},0,\ldots,0)\nonumber\\
&\qquad = T_m g(m-1,y_2,\ldots,y_{\numgold},0,\ldots,0) + (1-T_m) g(-(m-1),y_2,\ldots,y_{\numgold},0,\ldots,0)\label{eq:nec_confidence_5}
\end{align}
and
\begin{align}
S_m g(m,y_2,\ldots,y_{\numgold},0,\ldots,0) + (1-S_m)g(-m,y_2,\ldots,y_{\numgold},0,\ldots,0) = g(0,y_2,\ldots,y_{\numgold},0,\ldots,0)~. \label{eq:nec_confidence_5S}
\end{align}
We will now use this result in terms of function $g$ to get an equivalent result in terms of function $f$. For some $S \in \{0,\ldots,{\numgold}-1\}$, suppose $y_{{\numgold}-S+1}=0,\ldots,y_{\numgold}=0$. The remaining proof proceeds via an induction on $S$. We begin with $S={\numgold}-1$. In this case,~\eqref{eq:nec_confidence_5} and~\eqref{eq:nec_confidence_5S} simplify to
\begin{align}
&T_m g(m,0,\ldots,0) + (1-T_m)g(-m,0,0,\ldots,0)\nonumber\\
&\qquad = T_m g(m-1,0,\ldots,0) + (1-T_m) g(-(m-1),0,\ldots,0)
\end{align}
and
\begin{align}
S_m g(m,0,\ldots,0) + (1-S_m)g(-m,0,\ldots,0) = g(0,0,\ldots,0)~.
\end{align}
Applying the definition of function $g$ from~\eqref{eq:confidence_notation_subsample} leads to
\begin{align}
&T_m \left(c_1 f(m,0,\ldots,0)+c_2f(0,0,\ldots,0)\right) + (1-T_m) \left(c_1 f(-m,0,\ldots,0)+c_2f(0,0,\ldots,0)\right) \nonumber\\
&\quad= T_m \left(c_1 f(m-1,0,\ldots,0) +c_2f(0,0,\ldots,0)\right)  \nonumber \\
 &\qquad \qquad + (1-T_m) \left(c_1 f(-(m-1),0,\ldots,0)+c_2f(0,0,\ldots,0)\right),
\end{align}
and
\begin{align}
&S_m \left(c_1 f(m,0,\ldots,0)+c_2f(0,0,\ldots,0)\right) + (1-S_m) \left(c_1 f(-m,0,\ldots,0)+c_2f(0,0,\ldots,0)\right) \nonumber\\
&\qquad\qquad\qquad\qquad\qquad\qquad\qquad\qquad\qquad\qquad= \left(c_1 f(0,0,\ldots,0)+c_2f(0,0,\ldots,0)\right)
\end{align}
for constants $c_1>0$ and $c_2>0$ that respectively denote the probabilities that the first question is picked and not picked in the set of ${\numgold}$ gold standard questions. Cancelling out the common terms on both sides of the equation, we get the desired results
\begin{align}
T_m f(m,0,\ldots,0) & + (1-T_m)  f(-m,0,\ldots,0)  \nonumber \\
& = T_m f(m-1,0,\ldots,0) + (1-T_m) f(-(m-1),0,\ldots,0)
\end{align}
and
\begin{align}
&S_m f(m,0,\ldots,0) + (1-S_m) f(-m,0,\ldots,0) = f(0,0,\ldots,0)~.
\end{align}

Next, we consider the case of a general $S \in \{0,\ldots,\numgold-2\}$ and assume that the result is true when $y_{\numgold-S}=0,\ldots,y_\numgold=0$. In~\eqref{eq:nec_confidence_5} and~\eqref{eq:nec_confidence_5S}, the functions $g$ decompose into a sum of the constituent $f$ functions. These constituent functions $f$ are of two types: the first where all of the first $(\numgold - S)$ questions are included in the gold standard, and the second where one or more of the first $(\numgold - S)$ questions are not included in the gold standard. The second case corresponds to situations where there are more than $S$ questions skipped in the gold standard, i.e., when $y_{\numgold-S}=0,\ldots,y_\numgold=0$, and hence satisfies our induction hypothesis. The terms corresponding to these functions thus cancel out in the expansion of~\eqref{eq:nec_confidence_5} and~\eqref{eq:nec_confidence_5S}. The remainder comprises only evaluations of function $f$ for arguments in which the first $(\numgold-S)$ questions are included in the gold standard: since the last $(N-\numgold+S)$ questions are skipped by the worker, the remainder evaluates to
\begin{align*}
&T_m c_3 f(y_1,\ldots,y_{i-1},m,y_{i+1},\ldots,y_{\numgold}) + (1-T_m) c_3 f(y_1,\ldots,y_{i-1},-m,y_{i+1},\ldots,y_{\numgold})\nonumber\\
&\quad = T_m c_3  f(y_1,\ldots,y_{i-1},m-1,y_{i+1},\ldots,y_{\numgold}) + (1-T_m) c_3  f(y_1,\ldots,y_{i-1},-(m-1),y_{i+1},\ldots,y_{\numgold})~,\\
&S_m c_3  f(y_1,\ldots,y_{i-1},m,y_{i+1},\ldots,y_{\numgold}) + (1-S_m) c_3 f(y_1,\ldots,y_{i-1},-m,y_{i+1},\ldots,y_{\numgold})\nonumber\\
&\quad =  c_3 f(y_1,\ldots,y_{i-1},0,y_{i+1},\ldots,y_{\numgold})~,
\end{align*}
for some constant $c_3>0$. Dividing throughout by $c_3$ gives the desired result.

Finally, the arguments above hold for any permutation of the first ${\numgold}$ questions, thus completing the proof.

\subsection{Necessity of $T_\elll > S_\elll$ for the Problem to be Well Defined}
\label{app:SmorethanT}
We now show that the restriction $T_\elll > S_\elll$ was necessary when defining the thresholds in Section~\ref{sec:confidence}.

\begin{proposition}
Incentive-compatiblity necessitates $T_\elll > S_\elll~\forall~\elll \in \{2,\ldots,L\}$, even in the absence of the generalized-no-free-lunch axiom.
\label{prop:threshold_T_gt_S}
\end{proposition}

First observe that the proof of Lemma~\ref{lem:nec_confidence} did not employ the generalized-no-free-lunch axiom, neither did it assume $T_\elll > S_\elll$. We will thus use the result of Lemma~\ref{lem:nec_confidence} to prove our claim. 

Suppose the confidence of the worker for all but the first question is lower than $T_1$ and that the worker decides to skip all these questions. Suppose the worker attempts the first question. In order to ensure that the worker selects the answer that she believes is most likely to be true, it must be that
\begin{align}
f(\elll, 0,\ldots,0) > f(-\elll, 0,\ldots,0) \quad \forall \elll \in [L]~.
\end{align}

We now call upon Lemma~\ref{lem:nec_confidence} where we set $i=1$, $m = \elll$, $y_2=\ldots,y_\numgold=0$. Using the fact that $T_\elll > T_{\elll-1}~ \forall \elll \in \{2,\ldots,L\}$, we get
\begin{align}
T_\elll f(\elll, 0,\ldots,0) & + (1-T_\elll)f(-\elll, 0,\ldots,0) \nonumber \\ 
&= T_\elll f(\elll-1,0,\ldots,0) + (1-T_\elll) f(-(\elll-1),0,\ldots,0)\\
&> T_{\elll-1} f(\elll-1,0,\ldots,0) + (1-T_{\elll-1}) f(-(\elll-1),0,\ldots,0)\\
& =  T_{\elll-1} f(\elll-2,0,\ldots,0) + (1-T_{\elll-1}) f(-(\elll-2),0,\ldots,0)\\
& >T_{\elll-2} f(\elll-2,0,\ldots,0) + (1-T_{\elll-2}) f(-(\elll-2),0,\ldots,0)\\
& \vdots \nonumber\\
& >  T_{1} f(1,0,\ldots,0) + (1-T_{1}) f(-1,0,\ldots,0)\\
& = f(0,\ldots,0)\\
& = S_\elll f(\elll, 0,\ldots,0) + (1-S_\elll)f(-\elll, 0,\ldots,0).
\end{align}
Since $f(\elll, 0,\ldots,0)  > f(-\elll, 0,\ldots,0)$, we have our desired result.

\subsection{A Stronger No-free-lunch Condition: Impossibility Results}
\label{app:proof_strong_skip}
In this section, we prove the various claims regarding the strong no-free-lunch condition studied in Section~\ref{sec:strong_skip}.

\subsubsection{Proof of Proposition~\ref{prop:strong_unknowledgeable}}
If the worker skips all questions, then the expected payment is zero under the strong-no-free-lunch axiom. On the other hand, in order to incentivize knowledgeable workers to select answers whenever their confidences are greater than $T$, there must exist some situation in which the payment is strictly larger than zero. Suppose the payment is strictly positive when questions $\{1,\ldots,z\}$ are answered correctly, questions $\{z+1,\ldots,z'\}$ are answered incorrectly, and the remaining questions are skipped. If the confidence of the unknowledgeable worker is in the interval $(0,T)$ for every question, then attempting to answer questions $\{1,\ldots,z'\}$ and skipping the rest fetches her a payment that is strictly positive in expectation. Thus, this unknowledgeable worker is incentivized to answer at least one question.

\subsubsection{Proof of Proposition~\ref{prop:strong_knowledgeable}}
Consider a (knowledgeable) worker who has a confidence of $p \in (T,1]$ for the first question, $q \in (0,1)$ for the second question, and confidences in the interval $(0,T)$ for the remaining questions. Suppose the worker attempts to answer the first question (and selects the answer the believes is most likely to be correct) and skips the last $(\numques-2)$ questions as desired. Now, in order to incentivize her to answer the second question if $q>T$ and skip the second question if $q<T$, the payment mechanism must satisfy
\begin{align}
pq & g(1,1,0,\ldots,0) + (1-p) q g(-1,1,0,\ldots,0) + p(1-q) g(1,-1,0,\ldots,0) \nonumber \\
 & + (1-p)(1-q) g(-1,-1,0,\ldots,0) \overset{q<T}{\underset{q>T}{\lessgtr}}  p g(1,0,0,\ldots,0) + (1-p) g(-1,0,0,\ldots,0)~.
\end{align}
For any real-valued variable $q$, and for any real-valued constants $a$, $b$ and $c$, \[a q ~\overset{q<c}{\underset{q>c}{\lessgtr}}~b \quad \Rightarrow \quad \ ac=b~.\] As a result,
\begin{align}
p T & g(1,1,0,\ldots,0) + (1-p) T g(-1,1,0,\ldots,0) + p(1-T) g(1,-1,0,\ldots,0) \nonumber \\
& + (1-p)(1-T) g(-1,-1,0,\ldots,0) -  p g(1,0,0,\ldots,0) - (1-p) g(-1,0,0,\ldots,0) = 0~.
\end{align}
The left hand side of this equation is a polynomial in variable $p$ and takes a value of zero for all values of $p$ in a one-dimensional box $(T,1]$. It follows that the monomials of this polynomial must be zero, and in particular the constant term must be zero:
\begin{align}
T g(-1,1,0,\ldots,0) + (1-T) g(-1,-1,0,\ldots,0) - g(-1,0,0,\ldots,0) = 0~.
\end{align}
The strong-no-free-lunch condition implies $f(-1,-1,0,\ldots,0)=f(-1,0,\ldots,0) = f(0,\ldots,0) = 0$, and hence $g(-1,-1,0,\ldots,0)=g(-1,0,0,\ldots,0)=0$. Since $T \in (0,1)$, we have 
\begin{align}
0 &= g(-1,1,0,\ldots,0)\\
&= c_1 f(-1,1,0,\ldots,0) + c_2 f(-1,0,\ldots,0) + c_2 f(1,0,\ldots,0)~,
 \label{eq:strong_unique_proof_1}
\end{align}
for some constants $c_1>0$ and $c_2>0$ that represent the probability that the first two questions are included in the gold standard, and the probability that the first (or, second) but not the second (or, first) questions are included in the gold standard. Since $f$ is a non-negative function, it must be that
\begin{align}
f(1,0,\ldots,0)=0~.
\end{align}
Now suppose a (knowledgeable) worker has a confidence of $p \in (T,1]$ for the first question and confidences lower than $T$ for the remaining $(\numques-1)$ questions. Suppose the worker chooses to skip the last $(\numques-1)$ questions as desired. In order to incentivize the worker to answer the first question, the mechanism must satisfy for all $p \in (T,1]$,
\begin{align}
0 &< p g(1,0,\ldots,0) + (1-p) g(-1,0,\ldots,0) - g(0,0,\ldots,0)~\nonumber\\
& = p c_3 f(1,0,\ldots,0) + p c_4 f(0,0,\ldots,0) + (1-p) c_3 f(-1,0,\ldots,0) \nonumber \\
& \qquad \qquad + (1-p) c_4 f(0,0,\ldots,0) - f(0,0,\ldots,0)\nonumber\\
& = 0,
\end{align}
where $c_3>0$ and $c_4>0$ are some constants. The final equation is a result of the strong-no-free-lunch condition and the fact that $f(1,0,\ldots,0) =0$ as proved above. This yields a contradiction, and hence no incentive-compatible mechanism $f$ can satisfy the strong-no-free-lunch condition when $\numgold < \numques$ even when allowed to address only knowledgeable workers. 

Finally, as a sanity check, note that if ${\numgold}={\numques}$ then $c_2=0$ in~\eqref{eq:strong_unique_proof_1}. The proof above thus doesn't hold when ${\numgold}={\numques}$.

\subsubsection{Proof of Proposition~\ref{prop:mechanism_working_skip_strong}}
We will first show that the mechanism works as desired.

First consider the case when the worker is unknowledgeable and her confidences are of the form $T>p_{(1)}\geq p_{(2)}\geq p_{(3)} \geq \cdots \geq p_{({\numgold})}$. If she answers only the first question, then her expected payment is \[\kappa\frac{ p_{(1)}}{T}~.\] Let us now see her expected payment if she doesn't follow this answer pattern. The strong-no-free-lunch condition implies that if the worker doesn't answer any question then her expected payment is zero. Suppose the worker chooses to answer questions $\{i_1,\ldots,i_z\}$. In that case, her expected payment is 
\bea 
\kappa\frac{p_{i_1}\cdots p_{i_z}}{T^z}&=&\kappa\frac{p_{i_1}}{T}\cdots \frac{p_{i_z}}{T}\label{eq:mechanism_working_skip_strong_eq1}\\
&\leq &\kappa\left(\frac{p_{(1)}}{T}\right)^z\label{eq:mechanism_working_skip_strong_eq2}\\
&\leq & \kappa\frac{p_{(1)}}{T}~,\label{eq:mechanism_working_skip_strong_eq3}
\eea 
where~\eqref{eq:mechanism_working_skip_strong_eq3} uses the fact that $p_{(1)}<T$. The inequality in~\eqref{eq:mechanism_working_skip_strong_eq3} becomes an equality only when $z=1$. Now when $z=1$, the inequality in~\eqref{eq:mechanism_working_skip_strong_eq2} becomes an equality only when $i_1 = (1)$.  Thus the unknowledgeable worker is incentivized to answer only one question -- the one that she has the highest confidence in.

Now consider a knowledgeable worker and suppose her confidences are of the form $p_{(1)}\geq \cdots \geq p_{(m)} > T > p_{(m+1)} \geq \cdots \geq p_{({\numgold})}$ for some $m \geq 1$. If the worker answers questions $(1),\ldots,(m)$ as desired, her expected payment is \[\kappa\frac{p_{(1)}}{T}\cdots\frac{p_{(m)}}{T}~.\] Now let us see what happens if the worker does not follow this answer pattern. The strong-no-free-lunch condition implies that if the worker doesn't answer any question then her expected payment is zero.  Now, if she answers some other set of questions, say questions $\{i_1,\ldots,i_z\}$ with $p_{(1)} \leq p_{i_1}<\cdots<p_{i_y}\leq p_{(m)} < p_{i_{y+1}}<\cdots p_{i_z} \leq p_{({\numgold})}$. The expected payment in that case is 
\bea 
\kappa\frac{p_{i_1}\cdots p_{i_z}}{T^z}&=&\kappa\frac{p_{i_1}}{T}\cdots \frac{p_{i_z}}{T}\\&\leq &\kappa\frac{p_{i_1}}{T}\cdots \frac{p_{i_y}}{T}\label{eq:lem1_ineq1}\\
&\leq&\kappa\frac{p_{(1)}}{T}\cdots \frac{p_{(m)}}{T} \label{eq:lem1_ineq2}\eea
where inequality~\eqref{eq:lem1_ineq1} is a result of $\frac{p_{i_j}}{T} \leq 1~~\forall~j>y$ and holds with equality only when $y =z$. Inequality~\eqref{eq:lem1_ineq2} is a result of $\frac{p_{(j)}}{T} \geq 1~~\forall~j\leq m$ and holds with equality only when $y = m$. Thus the expected payment is maximized when $i_1=(1),\ldots,i_z=(m)$ as desired. Finally, the payment strictly increases with an increase in the confidences, and hence the worker is incentivized to always consider the answer that she believes is most likely to be correct.

We will now show that this mechanism is unique.

The necessary conditions derived in Lemma~\ref{lem:nec_skip}, when restricted to $\numgold=\numques$ and $(y_1,\ldots,y_{i-1},y_{i+1},\ldots,y_\numgold) \neq \{0\}^{\numques-1}$, is also applicable to the present setting. This is because the strong-no-free-lunch condition assumed here is a stronger condition than the no-free-lunch axiom considered in Lemma~\ref{lem:nec_skip}, and moreover, $(y_1,\ldots,y_{i-1},y_{i+1},\ldots,y_\numgold) \neq \{0\}^{\numques-1}$ avoids the use of unknowledgeable workers in the proof of Lemma~\ref{lem:nec_skip}. It follows that for every question $i \!\in\! \{1,\ldots,{\numgold}\}$ and every $(y_1,\ldots,y_{i-1},y_{i+1},\ldots,y_{\numgold} )\! \in\! \{-1,0,1\}^{{\numgold}-1} \backslash \{0\}^{\numgold-1}$, it must be that
\begin{align}
&T f(y_1,\ldots,y_{i-1},1,y_{i+1},\ldots,y_{\numgold}) + (1-T) f(y_1,\ldots,y_{i-1},-1,y_{i+1},\ldots,y_{\numgold})\nonumber\\
&\qquad\qquad\qquad\qquad\qquad\qquad\qquad\qquad\qquad\qquad = f(y_1,\ldots,y_{i-1},0,y_{i+1},\ldots,y_{\numgold})~.\label{eq:lineareqn_strong_skip}
\end{align}

We claim that the payment must be zero whenever the number of incorrect answers $W>0$. The proof proceeds by induction on the number of correct answers $C$. First suppose $C=0$ (and $W>0$). Then all questions are either wrong or skipped, and hence by the strong-no-free-lunch condition, the payment must be zero. Now suppose the payment is necessarily zero whenever $W>0$ and the total number of correct answers is $(C-1)$ or lower, for some $C \in [\numgold-1]$. Consider any  evaluation $(y_1,\ldots,y_\numgold) \in \{-1,0,1\}^{\numgold}$ in which the number of incorrect answers is more than zero and the number of correct answers is $C$. Suppose $y_i=1$ for some $i \in [\numgold]$, and $y_j=-1$ for some $j \in [\numgold] \backslash \{i\}$. Then from the induction hypothesis, we have $f(y_1,\ldots,y_{i-1},-1,y_{i+1},\ldots,y_{\numgold}) =f(y_1,\ldots,y_{i-1},0,y_{i+1},\ldots,y_{\numgold}) = 0$. Applying~\eqref{eq:lineareqn_strong_skip} and noting that $T \in (0,1)$, we get that $f(y_1,\ldots,y_{i-1},1,y_{i+1},\ldots,y_{\numgold})=0$ as claimed. This result also allows us to simplify~\eqref{eq:lineareqn_strong_skip} to: For every question $i \in \{1,\ldots,{\numgold}\}$ and every $(y_1,\ldots,y_{i-1},y_{i+1},\ldots,y_{\numgold} ) \in \{-1,0,1\}^{{\numgold}-1} \backslash \{0\}^{\numgold-1}$,
\begin{align}
f(y_1,\ldots,y_{i-1},1,y_{i+1},\ldots,y_{\numgold}) = \frac{1}{T} f(y_1,\ldots,y_{i-1},0,y_{i+1},\ldots,y_{\numgold})~.\label{eq:lineareqn_strong2_skip}
\end{align}

We now show that when $C>0$ and $W=0$, the payment must necessarily be of the form described in the statement of Proposition~\ref{prop:mechanism_working_skip_strong}. The proof again proceeds via an induction on the number of correct answers $C~(\geq 1)$. Define a quantity $\kappa >0$ as
\begin{align}
\kappa = T f(1,0,\ldots,0)~. \label{eq:kappa_strong_proof}
\end{align}

Now consider the payment $f(1,y_2,\ldots,y_\numgold)$ for some $(y_2,\ldots,y_\numgold) \in \{0,1\}^{\numgold-1}\backslash \{0\}^{\numgold-1}$ with $C$ correct answers. Applying~\eqref{eq:lineareqn_strong2_skip} repeatedly (once for every $i$ such that $y_i=1$), we get 
\begin{align}
f(1,y_2,\ldots,y_\numgold) = \frac{\kappa}{T^C}~.
\end{align}

Unlike other results in this paper, at this point we cannot claim the result to hold for all permutations of the questions. This is because we have defined the quantity $\kappa$ in an asymmetric manner~\eqref{eq:kappa_strong_proof}, in terms of the payment function when the {\it first} question is correct and the rest are skipped. In what follows, we will prove that the result claimed in the statement of Proposition~\ref{prop:mechanism_working_skip_strong} indeed holds for all permutations of the questions. 

From~\eqref{eq:lineareqn_strong2_skip} we have
\begin{align}
f(0,1,0,\ldots,0) &= T f(1,1,0,\ldots,0)\\
& = f(1,0,0,\ldots,0)\\
& = \kappa~.
\end{align}
Thus the payment must be $\kappa$ even if the second answer in the gold standard is correct and the rest are skipped. In fact, the argument holds when any one answer in the gold standard is correct and the rest are skipped. Thus the definition of $\kappa$ is not restricted to the first question alone as originally defined in~\eqref{eq:kappa_strong_proof}, but holds for all permutations of the questions. This allows the other arguments above to be applicable to any permutation of the questions. Finally, the budget constraint of $\maxpay$ fixes the value of $\kappa$ to that claimed, thereby completing the proof.

\subsubsection{Proof of Proposition~\ref{prop:strong_confidence}}
Proposition~\ref{prop:mechanism_working_skip_strong} proved that under the skip-based setting with the strong-no-free-lunch condition, the payment must be zero when one or more answers are incorrect. This part of the proof of Proposition~\ref{prop:mechanism_working_skip_strong} holds even when $L>1$. It follows that for any question, the penalty for an incorrect answer is the same for any confidence-level in $\{1,\ldots,L\}$. Thus the worker is incentivized to always select that confidence-level for which the payment is the maximum when the answer is correct, irrespective of her own confidence about the question. This contradicts our requirements.

\section{Details of Experiments}\label{app:experiments}
In this section, we provide further details about the experiments  described earlier in Section~\ref{sec:experiments}. The experiments were carried out on the Amazon Mechanical Turk (\url{mturk.com}) online crowdsourcing platform in the time period June to October 2013. Figure~\ref{fig:interface_supplement} illustrates the interface shown to the workers for each of the experiments described in Section~\ref{sec:experiments}, while Figure~\ref{fig:experiments_instructions_supplement} depicts the instructions given to the workers. 
The following are more details of each individual experiment. In the description, the notation $\kappa$ is as defined in Algorithm~\ref{algo:incentive_skip} and Algorithm~\ref{algo:incentive_confidence}, namely, $\kappa = (\maxpay-\minpay) T^\numgold$ for the skip-based setting and $\kappa = (\maxpay - \minpay) \left(\frac{1}{\alpha_L}\right)^{\numgold}$ for the confidence-based setting.

\subsection{Recognizing the Golden Gate Bridge}
\begin{figure}[h!]
\centering
\includegraphics[width=\textwidth]{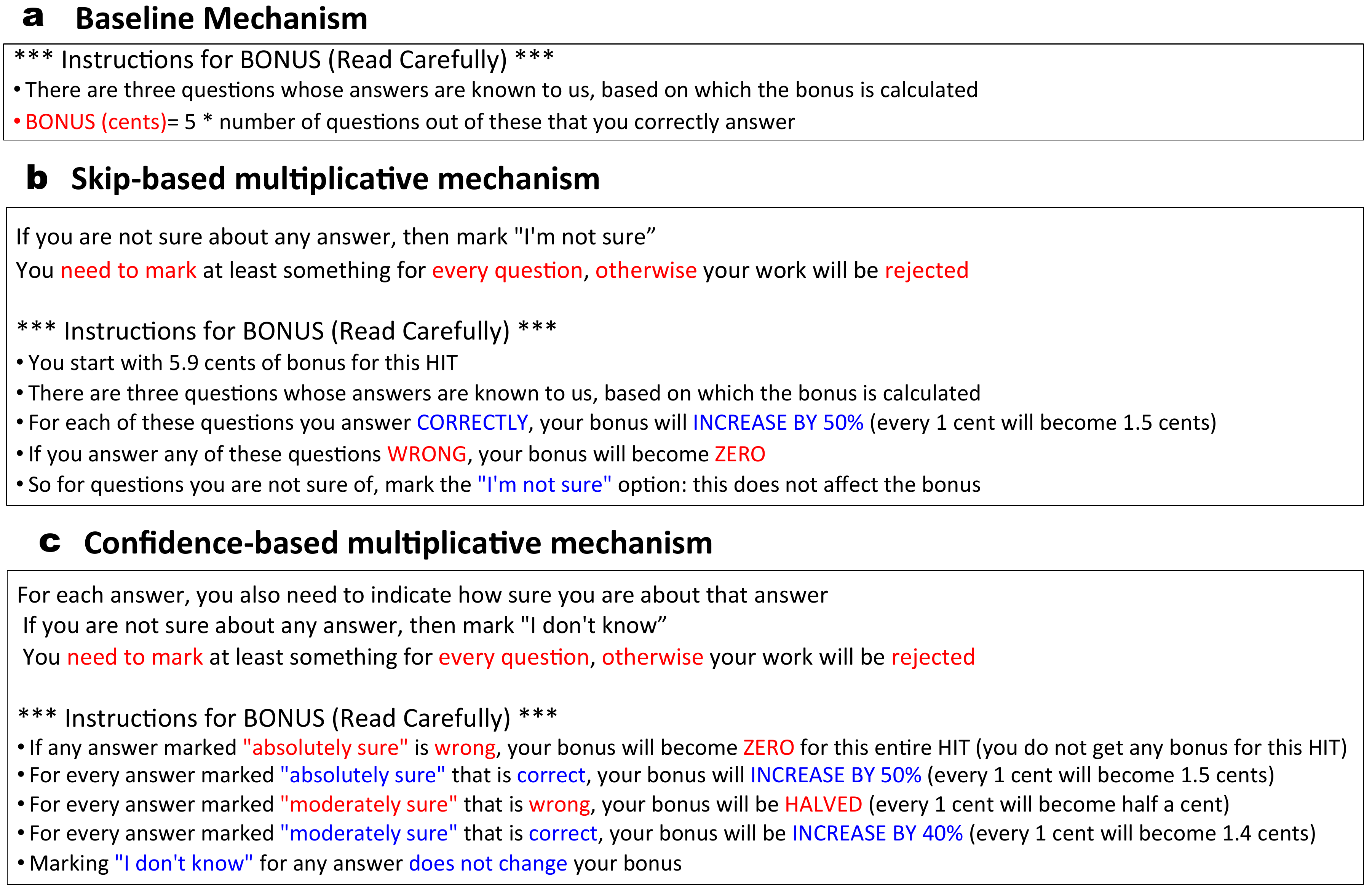}
\caption{An example of the instructions displayed to the worker under the three mechanisms.}
\label{fig:experiments_instructions_supplement}
\end{figure}

A set of $21$ photographs of bridges were shown to the workers, and for each photograph, they had to identify if it depicted the Golden Gate Bridge or not. An example of this task is depicted in Figure~\ref{fig:interface_supplement}{a}, and the instructions provided to the worker under the three mechanisms are depicted in Figure~\ref{fig:experiments_instructions_supplement}. The fixed amount offered to workers was $\minpay=3$ cents for the task, and the bonus was based on $3$ gold standard questions. We compared (a) the baseline mechanism with $5$ cents for each correct answer in the gold standard, (b) the skip-based mechanism with $\kappa=5.9$ and $\frac{1}{T}=1.5$, and (c) the confidence-based mechanism with $\kappa=5.9$ cents, $L=2$, $\alpha_2=1.5,\ \alpha_1=1.4,\ \alpha_0=1,\ \alpha_{-1}=0.5,\ \alpha_{-2}=0$. The results of this experiment are presented in Figure~\ref{fig:plots_supplement}{a}.
 
\subsection{Transcribing Vehicles' License Plate Numbers from Photographs}
This task presented the workers with $18$ photographs of cars and asked them to transcribe the license plate numbers from each of them (source of photographs: \textrm{http://www.coolpl8z.com}). An example of this task is depicted in Figure~\ref{fig:interface_supplement}{b}. The fixed amount offered to workers was $\minpay=4$ cents for the task, and the bonus was based on $4$ gold standard questions. We compared (a) the baseline mechanism with $10$ cents for each correct answer in the gold standard, (b) the skip-based mechanism with $\kappa=0.62$ and $\frac{1}{T}=3$, and (c) the confidence-based mechanism with $\kappa=3.1$ cents, $L=2$, $\alpha_2=2,\ \alpha_1=1.95,\ \alpha_0=1,\ \alpha_{-1}=0.5\ \alpha_{-2}=0$. The results of this experiment are presented in Figure~\ref{fig:plots_supplement}{b}. When evaluating, in the worker's answers as well as in the true solutions, we converted all text to upper case, and removed all spaces and punctuations. We then declared a worker's answer to be in error if it did not have an exact match with the true solution.

\subsection{Classifying Breeds of Dogs}
This task required workers to identify the breeds of dogs shown in 85 images (source of images:~\citetn{khosla2011novel,deng2009imagenet}). For each image, the worker was given ten breeds to choose from. An example of this task is depicted in Figure~\ref{fig:interface_supplement}{c}. The fixed amount offered to workers was $\minpay=5$ cents for the task, and the bonus was based on $7$ gold standard questions. We compared (a) the baseline mechanism with $8$ cents for each correct answer in the gold standard, (b) the skip-based mechanism with $\kappa=0.78$ and $\frac{1}{T}=2$, and (c) the confidence-based mechanism with $\kappa=0.78$ cents, $L=2$, $\alpha_2=2,\ \alpha_1=1.66,\ \alpha_0=1,\ \alpha_{-1}=0.67,\ \alpha_{-2}=0$. The results of this experiment are presented in Figure~\ref{fig:plots_supplement}{c}.

\subsection{Identifying Heads of Countries}
Names of $20$ personalities were provided and had to be classified as to whether they were ever the (a) President of the USA, (b) President of India, (c) Prime Minister of Canada, or (d) neither of these. An example of this task is depicted in Figure~\ref{fig:interface_supplement}{d}. The fixed amount offered to workers was $\minpay=2$ cents for the task, and the bonus was based on $4$ gold standard questions. While the ground truth in most other multiple-choice experiments had approximately an equal representation from all classes, this experiment was heavily biased with one of the classes never being correct and another being correct for just $3$ of the $20$ questions. We compared (a) the baseline mechanism with $2.5$ cents for each correct answer in the gold standard, (b) the skip-based mechanism with $\kappa=0.25$ and $\frac{1}{T}=3$, and (c) the confidence-based mechanism with $\kappa=1.3$ cents, $L=2$, $\alpha_2=2,\ \alpha_1=1.95,\ \alpha_0=1,\ \alpha_{-1}=0.5,\ \alpha_{-2}=0$. The results of this experiment are presented in Figure~\ref{fig:plots_supplement}{d}.

\subsection{Identifying Flags}
This was a relatively long task, with 126 questions. Each question required the workers to identify if a displayed flag belonged to a place in (a) Africa, (b) Asia/Oceania, (c) Europe, or (d) neither of these. An example of this task is depicted in Figure~\ref{fig:interface_supplement}{e}. The fixed amount offered to workers was $\minpay=4$ cents for the task, and the bonus was based on $8$ gold standard questions. We compared (a) the baseline mechanism with $4$ cents for each correct answer in the gold standard, (b) the skip-based mechanism with $\kappa=0.2$ and $\frac{1}{T}=2$, and (c) the confidence-based mechanism with $\kappa=0.2$ cents, $L=2$, $\alpha_2=2,\ \alpha_1=1.66,\ \alpha_0=1,\ \alpha_{-1}=0.67,\ \alpha_{-2}=0$. The results of this experiment are presented in Figure~\ref{fig:plots_supplement}{e}.

\subsection{Distinguishing Textures}
This task required the workers to identify the textures shown in 24 grayscale images (source of images:~\citetn[Dataset 1: Textured surfaces]{lazebnik2005sparse}). For each image, the worker had to choose from $8$ different options. Such a task has applications in computer vision, where it aids in recognition of objects or their surroundings. An example of this task is depicted in Figure~\ref{fig:interface_supplement}{f}. The fixed amount offered to workers was $\minpay=3$ cents for the task, and the bonus was based on $4$ gold standard questions. We compared (a) the baseline mechanism with $10$ cents for each correct answer in the gold standard, (b) the skip-based mechanism with $\kappa=3.1$ and $\frac{1}{T}=2$, and (c) the confidence-based mechanism with $\kappa=3.1$ cents, $L=2$, $\alpha_2=2,\ \alpha_1=1.66,\ \alpha_0=1,\ \alpha_{-1}=0.67,\ \alpha_{-2}=0$. The results of this experiment are presented in Figure~\ref{fig:plots_supplement}{f}.

\subsection{Transcribing Text from an Image: Film Certificate}
The task showed an image containing $11$ (short) lines of blurry text which the workers had to decipher. We used text from a certain certificate which movies releasing in India are provided. We slightly modified its text in order to prevent workers from searching a part of it online and obtaining the entire text by searching the first few transcribed lines on the internet.  An example of this task is depicted in Figure~\ref{fig:interface_supplement}{g}. The fixed amount offered to workers was $\minpay=5$ cents for the task, and the bonus was based on $2$ gold standard questions. We compared (a) the baseline mechanism with $20$ cents for each correct answer in the gold standard, (b) the skip-based mechanism with $\kappa=5.5$ and $\frac{1}{T}=3$, and (c) the confidence-based mechanism with $\kappa=12.5$ cents, $L=2$, $\alpha_2=2,\ \alpha_1=1.95,\ \alpha_0=1,\ \alpha_{-1}=0.5,\ \alpha_{-2}=0$. The results of this experiment are presented in Figure~\ref{fig:plots_supplement}{g}. When evaluating, in the worker's answers as well as in the true solutions, we converted all text to upper case, and removed all spaces and punctuations. We then declared a worker's answer to be in error if it did not have an exact match with the true solution.

\subsection{Transcribing Text from an Image: Script of a Play}
The task showed an image containing $12$ (short) lines of blurry text which the workers had to decipher. We borrowed a paragraph from Shakespeare's play `As You Like It.' We slightly modified the text of the play in order to prevent workers from searching a part of it online and obtaining the entire text by searching the first few transcribed lines on the internet. An example of this task is depicted in Figure~\ref{fig:interface_supplement}{h}.  The fixed amount offered to workers was $5$ cents for the task, and the bonus was based on $2$ gold standard questions. We compared (a) the baseline mechanism with $\minpay=20$ cents for each correct answer in the gold standard, (b) the skip-based mechanism with $\kappa=5.5$ and $\frac{1}{T}=3$, and (c) the confidence-based mechanism with $\kappa=12.5$ cents, $L=2$, $\alpha_2=2,\ \alpha_1=1.95,\ \alpha_0=1,\ \alpha_{-1}=0.5,\ \alpha_{-2}=0$. The results of this experiment are presented in Figure~\ref{fig:plots_supplement}{h}. When evaluating, in the worker's answers as well as in the true solutions, we converted all text to upper case, and removed all spaces and punctuations. We then declared a worker's answer to be in error if it did not have an exact match with the true solution.

\subsection{Transcribing Text from Audio Clips}
\begin{figure}[h!]
\centering
\includegraphics[width=.8\textwidth]{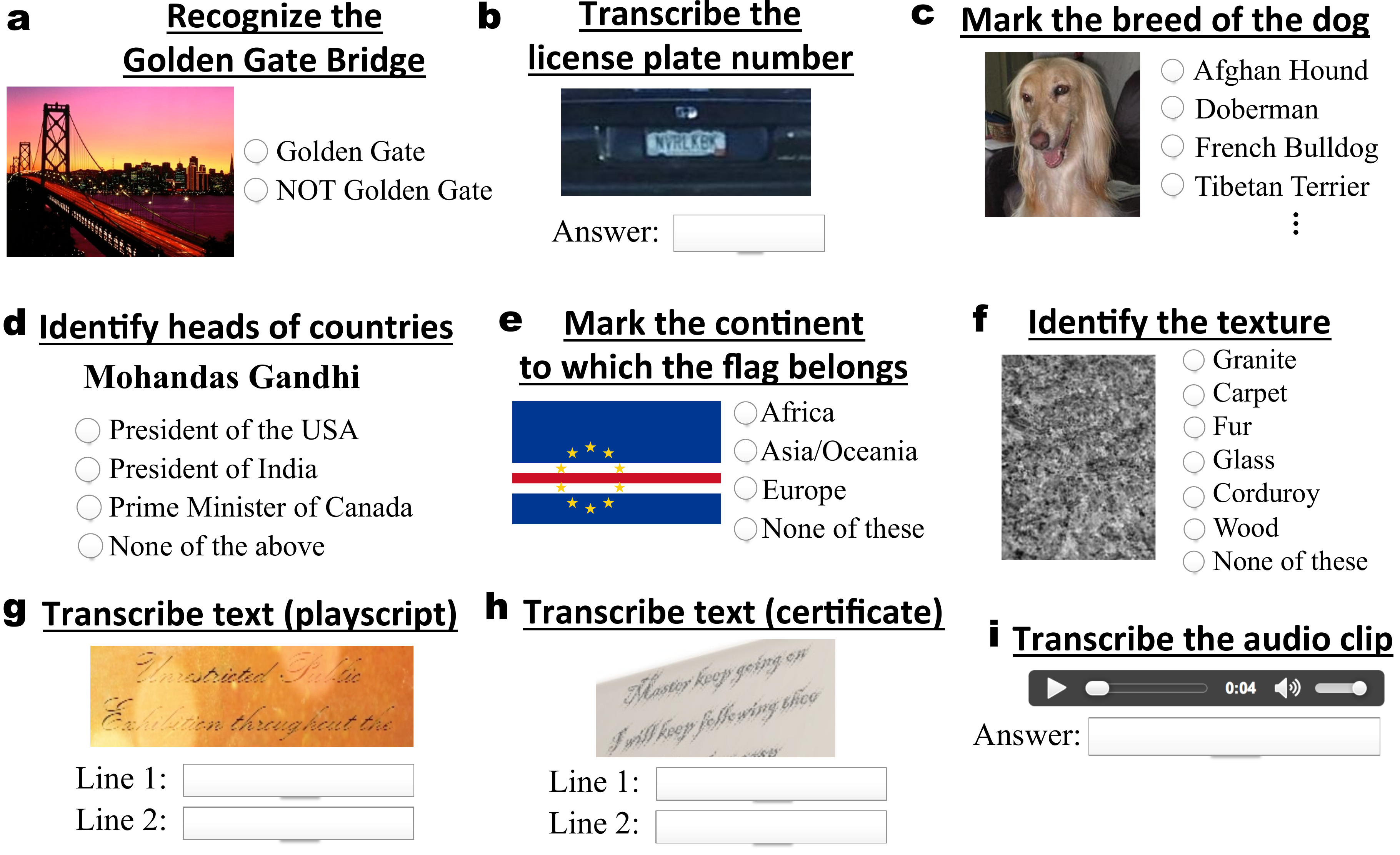}
\caption{Various tasks on which the payment mechanisms were tested. The interfaces shown are that of the baseline mechanism, i.e., without the skipping or confidence choices.}
\label{fig:interface_supplement}
\end{figure}

The workers were given $10$ audio clips which they had to transcribe to text. Each audio clip was $3$ to $6$ seconds long, and comprised of a short sentence, e.g., ``my favourite topics of conversation are sports, politics, and movies.'' Each of the clips were recorded in different accents using a text-to-speech converter. An example of this task is depicted in Figure~\ref{fig:interface_supplement}{i}. The fixed amount offered to workers was $\minpay=5$ cents for the task, and the bonus was based on $2$ gold standard questions. We compared (a) the baseline mechanism with $20$ cents for each correct answer in the gold standard, (b) the skip-based mechanism with $\kappa=5.5$ and $\frac{1}{T}=3$, and (c) the confidence-based mechanism with $\kappa=12.5$ cents, $L=2$, $\alpha_2=2,\ \alpha_1=1.95,\ \alpha_0=1,\ \alpha_{-1}=0.5,\ \alpha_{-2}=0$. The results of this experiment are presented in Figure~\ref{fig:plots_supplement}{i}.

\vspace{-.6cm}

\section{General Utility Functions}\label{sec:utility}
\newcommand{\utilityfn}{U}
\newcommand{\utilitycodomain}{\mathcal{I}}
In this section, we consider a setting where the worker, instead of maximizing her expected payment, aims to maximize the expected value of some \textit{utility function} of her payment. Consider any function $\utilityfn: \mathbb{R}_+ \rightarrow \utilitycodomain$, where $\utilitycodomain$ is any interval on the real number line. We will require the function $\utilityfn$ to be strictly increasing and to have an inverse. Examples of such functions include $\utilityfn(x) = \log(1+x)$ with $\utilitycodomain = \mathbb{R}_+$, $\utilityfn(x) = \sqrt{x}$ with $\utilitycodomain = \mathbb{R}_+$, and $\utilityfn(x) = 1-e^{-x}$ with $\utilitycodomain = [0,1]$. For any payment $f$ made to the worker (based on the evaluation of her answers to the gold standard questions), her utility for this payment is $\utilityfn(f)$. The worker aims to maximize the expected value of $\utilityfn(f)$, where the expectation is with respect to her beliefs regarding correctness of her answers and the uniformly random distribution of the $\numgold$ gold standard questions among the set of $\numques$ questions. The function $\utilityfn$ is assumed to be known to the worker as well as the system designer.

Consider the confidence-based setting of Section~\ref{sec:confidence} (of which, the skip-based setting of Section~\ref{sec:skip} is a special case). Recall the notation $\{\ans{i}\}_{i=1}^{\numgold}$, $\{\alpha_j\}_{j=-L}^{L}$ and $\kappa$ from Algorithm~\ref{algo:incentive_confidence}. Also recall the (generalized-)no-free-lunch axiom which mandates a zero payment if, in the gold standard, (all attempted questions are marked as the highest confidence $L$ and) the answers to all the attempted questions are incorrect. The following proposition extends the results of the main text in the paper to this setting with utility functions.

\begin{proposition}\label{thm:utility}
For a worker who aims to maximize function $\utilityfn$ of the payment, the one and only mechanism that is incentive-compatible and satisfies the (generalized-)no-free-lunch axiom is
\[
\textrm{Payment}(\ans{1},\ldots,\ans{\numgold}) = U^{-1}\left(\kappa \prod_{i=1}^{\numgold} \alpha_{\ans{i}} + U(\minpay)\right)~,
\]
where the constants $\{\alpha_j\}_{j=-L}^{L}$ are as defined in Algorithm~\ref{algo:incentive_confidence} and $\kappa = (\utilityfn(\maxpay) - \utilityfn(\minpay)) \alpha_L^{-\numgold}$.
\end{proposition}
\noindent Note that for the problem to be well defined, the interval $[\minpay, \maxpay]$ should be contained in the interval $\utilitycodomain$. The proof of Proposition~\ref{thm:utility} follows easily from the results proved earlier in the paper, and is provided below for completeness.


\paragraph{Proof of Proposition~\ref{thm:utility}.} 
We will first verify that the proposed payment is always non-negative and satisfies the (generalized-)no-free-lunch axiom. Recall from Theorem~\ref{thm:mechanism_working_confidence} that for every evaluation $\{\ans{1},\ldots,\ans{\numgold}\}$ for which the (generalized-)no-free-lunch axiom mandates a zero payment, the value of $\kappa \prod_{i=1}^{\numgold} \alpha_{\ans{i}}$ is zero. It follows that the payment $U^{-1}\left(\kappa \prod_{i=1}^{\numgold} \alpha_{\ans{i}} + U(\minpay)\right)=U^{-1}(0+U(\minpay))=\minpay$, where the final equation is a consequence of the invertibility of $U$. Further, recall that the value of $\kappa \prod_{i=1}^{\numgold} \alpha_{\ans{i}}$ in Algorithm~\ref{algo:incentive_confidence} is never smaller than zero. Since the function $U$ is increasing, so is $U^{-1}$, and hence the payment is always non-negative.

We will now prove that the proposed payment is incentive-compatible. To this end, observe that the utility of the proposed payment is 
\begin{align*}
\utilityfn(\textrm{Payment}) &= \utilityfn\left(\utilityfn^{-1}\left(\kappa \prod_{i=1}^{\numgold} \alpha_{\ans{i}} + U(\minpay)\right)\right) \\
& = \kappa \prod_{i=1}^{\numgold} \alpha_{\ans{i}} + \utilityfn(\minpay)~.
\end{align*}
Noting that $\utilityfn(0)$ is a constant independent of the worker's answers, the result of Theorem~\ref{thm:mechanism_working_confidence} implies that the expectation of $\utilityfn({\rm Payment})$ behaves exactly as required for incentive-compatibility.

We will now prove uniqueness of this mechanism. Replacing $\payfn(\cdot)$ by $\utilityfn(\textrm{Payment}(\cdot))$ in the proof of Theorem~\ref{thm:unique_confidence}, we get that the function $\utilityfn(\textrm{Payment})$ must be of the form
\begin{align*}
U(\textrm{Payment}(\ans{1},\ldots,\ans{\numgold})) = c_1 \prod_{i=1}^{\numgold} \alpha_{\ans{i}}  + c_2,
\end{align*}
for some constants $c_1$ and $c_2$, where $\{\alpha_{\ans{j}}\}_{j=-L}^{L}$  are as defined in Algorithm~\ref{algo:incentive_confidence}. In other words, the payment must be of the form
\begin{align*}
\textrm{Payment}(\ans{1},\ldots,\ans{\numgold}) = \utilityfn^{-1}\left(c_1 \prod_{i=1}^{\numgold} \alpha_{\ans{i}}  + c_2 \right).
\end{align*}
One can evaluate that the maximum value of this payment is $c_1 + c_2$. From our $\maxpay$-budget constraint, we then have $c_1 + c_2 = \maxpay$. Furthermore, When the evaluations $\ans{1},\ldots,\ans{\numgold}$ are such that the (generalized-)no-free-lunch applies, we need $\textrm{Payment}=\minpay$. It follows that $c_2 =\utilityfn(\minpay)$, and consequently $c_1 = \utilityfn(\maxpay) - \utilityfn(\minpay)$, thereby completing the proof.

\ifjmlr
    \bibliography{crowdsourcing}
\fi

\end{document}